\documentclass{article}[11pt]
\usepackage{jheppub}
\pdfoutput=1
\usepackage{graphicx}
\usepackage{amsmath,amssymb,mathrsfs}
\usepackage{bbm}
\usepackage{color}
\usepackage{xcolor}
\usepackage{dsfont}
\usepackage{cancel}
\usepackage{dsfont}
\usepackage{epstopdf}
\usepackage{epsfig}
\usepackage{bm}
\usepackage{dcolumn}
\usepackage{hyperref}
\usepackage{enumitem}
\usepackage{multirow}
\usepackage{lineno}
\usepackage{mathtools,slashed}
\usepackage{url}

\newcommand{\ben}{\begin{enumerate}}
\newcommand{\een}{\end{enumerate}}
\newcommand{\bit}{\begin{itemize}}
\newcommand{\eit}{\end{itemize}}

\newcommand{\beqa}{\begin{eqnarray}}
\newcommand{\eeqa}{\end{eqnarray}}
\newcommand{\beq}{\begin{equation}}
\newcommand{\eeq}{\end{equation}}
\newcommand{\bay}{\begin{array}}
\newcommand{\eay}{\end{array}}

\def\ifmath#1{\relax\ifmmode #1\else $#1$\fi}

\newcommand{\mg}{{\sc MG5\_}a{\sc MC@NLO}}
\newcommand{\mgvtwo}{{\sc MG5\_}a{\sc MC@NLO}v2.4.2}

\def\gsim{\ \rlap{\raise 3pt \hbox{$>$}}{\lower 3pt \hbox{$\sim$}}\ }
\def\lsim{\ \rlap{\raise 3pt \hbox{$<$}}{\lower 3pt \hbox{$\sim$}}\ }

\def\ls#1{\ifmath{_{\lower1.5pt\hbox{$\scriptstyle #1$}}}}
\def\lsup#1{^{\lower 6pt\hbox{$\scriptstyle#1$}}}

\def\squark{\tilde q}
\def\ninoone{{\tilde \chi}^0_1}

\def\Tr{{{\rm Tr}}}

\def\bracket#1#2 {\mathinner{\langle{#1}|{#2}\rangle}}

\def\bracket#1#2 {\mathinner{\langle{#1}|{#2}\rangle}}

\newcommand{\MET}{\slashed{E}_T}

\newcommand{\be}{\begin{equation}}
\newcommand{\ee}{\end{equation}}
\newcommand{\bea}{\begin{eqnarray}}
\newcommand{\eea}{\end{eqnarray}}

\graphicspath{{figs/}}
\begin{document}


\title{Simplified Models for Displaced Dark Matter Signatures}

\author[a]{Oliver Buchmueller}
\emailAdd{oliver.buchmueller@cern.ch}
\affiliation[a]{High Energy Physics Group, Blackett Laboratory, Imperial College, Prince Consort Road, London, SW7 2AZ, UK}

\author[b,c]{, Albert De Roeck}
\emailAdd{albert.de.roeck@cern.ch}
\affiliation[b]{Physics Department, CERN, CH-1211 Gen\`eve 23, Switzerland}
\affiliation[c]{Antwerp University, B2610 Wilrijk, Belgium}

\author[d]{, Matthew McCullough}
\emailAdd{matthew.mccullough@cern.ch}
\affiliation[d]{Theoretical Physics Department, CERN, CH-1211 Gen\`eve 23, Switzerland}

\author[e]{, Kristian~Hahn}
\emailAdd{kristian.hahn@cern.ch}
\affiliation[e]{Department of Physics and Astronomy, Northwestern University, Evanston, Illinois 60208, USA}

\author[e]{, Kevin Sung}
\emailAdd{kevin.kai.hong.sung@cern.ch}

\author[f]{, Pedro Schwaller}
\emailAdd{pedro.schwaller@uni-mainz.de}
\affiliation[f]{PRISMA Cluster of Excellence \& Institute of Physics, Johannes Gutenberg University, 55099 Mainz, Germany}

\author[d]{, Tien-Tien Yu}
\emailAdd{tien-tien.yu@cern.ch}

\preprint{MITP/17-025, CERN-TH-2017-091}

\abstract{
We propose a systematic programme to search for long-lived neutral particle signatures through a minimal set of displaced $\MET$ searches (dMETs).
 Our approach is to extend the well-established dark matter simplified models to include displaced vertices.  The dark matter simplified models are used to describe the primary production vertex.  A displaced secondary vertex, characterised by the mass of the long-lived particle and its lifetime, is added for the displaced signature.  We show how these models can be motivated by, and mapped onto, complete models such as gauge-mediated SUSY breaking and models of neutral naturalness.  We also outline how this approach may be used to extend other simplified models to incorporate displaced signatures and to characterise searches for long-lived charged particles.  Displaced vertices are a striking signature with virtually no backgrounds from SM processes, and thus provide an excellent target for the high-luminosity run of the Large Hadron Collider. The proposed models and searches provide a first step towards a systematic broadening of the displaced dark matter search programme. 
}
\maketitle
\section{Introduction} 
The nature of dark matter (DM) remains a mystery. One explanation is that dark matter is a new, beyond the Standard Model (SM) particle. The most popular paradigm is that of a single weakly interacting massive particle (WIMP) with weak scale mass and coupling to the SM. However, there is also the possibility that the dark sector possesses a richer structure, replete with additional particles, some of which may be long-lived. Indeed, many extensions of the SM predict the existence of new particles with long lifetimes (see {\emph {e.g.}}~\cite{Strassler:2006im,Fairbairn:2006gg,Pospelov:2007mp}). These particles can be produced in collider experiments and can propagate large distances in the detector before they decay.  

While common reconstruction algorithms assume prompt decays of particles at the production vertex, long-lived particles can lead to a variety of unique signatures.  Depending on the particle nature and its lifetime, these signatures can include displaced vertices, disappearing tracks, massive particle tracks, jet production outside of collision event windows, and the production of collimated jets of leptons, amongst other possibilities~\cite{Strassler:2006im,Strassler:2006ri,Cheung:2009su,Aad:2014yea,Aad:2012kw,Meade:2009mu,Falkowski:2010cm,Meade:2011du,Meade:2010ji,Jaiswal:2013xra,Buckley:2014ika,Cui:2014twa,Helo:2013esa,Schwaller:2015gea,Liu:2015bma,Clarke:2015ala,Buschmann:2015awa,Curtin:2015fna,Csaki:2015fba,Coccaro:2016lnz}. Each of these signatures requires the development of dedicated reconstruction and selection techniques to ensure their efficient inclusion in high-level physics analysis. Furthermore, these exotic signatures are produced from relatively unique processes, distinct from backgrounds, and make for an excellent target for the high-luminosity run of the Large Hadron Collider (LHC). It is therefore timely to develop a broad programme to search for rich dark sectors.

Although the current LHC experiments presently set strong limits on models of Supersymmetry (SUSY) that can predict new long-lived particles (see, for example,~\cite{Khachatryan:2010uf,CMS:2014wda,Chatrchyan:2013oca,CMS:2014hka,Khachatryan:2016sfv,Aad:2015oga,Aad:2015uaa,Aad:2015asa,Aad:2014yea,ATLAS:2012av}), displaced signatures for other models are either only partially covered or not covered at all.  The ATLAS and CMS experiments have executed only a few dedicated analyses that combine long-lived signatures with topologies that include significant missing transverse energy ($\MET$), for instance.  The most prominent examples of such analyses are the searches for displaced photons with significant $\MET$ ~\cite{Aad:2014gfa,Chatrchyan:2012ir} that are interpreted in the framework of Gauge Mediated Supersymmetry Breaking (GMSB), and the searches for disappearing tracks~\cite{CMS:2014gxa,Aad:2013yna} that are motivated by long-lived chargino production in context of the Anomaly-Mediated Supersymmetry Breaking (AMSB) model. One of the main challenges in improving the coverage of long-lived signatures is related to the means by which the relevant parameter space and corresponding collider signatures of long-lived particles are currently explored. At present, the main option for long-lived $\MET$ searches is to re-interpret the canonical $\MET$ analyses, which are mainly used to search for SUSY production in proton-proton collisions.  While the SUSY search programme at the LHC spans across a large variety of different final states, including multi-jet, multi-lepton/photon and $\MET$ topologies, the $\MET$ signature is assumed to originate from prompt decays.  The traditional $\MET$ searches are therefore typically far from being optimal for long-lived scenarios. Furthermore, as for example outlined in~\cite{Allanach:2016pam}, the recasting of these searches in the context of long-lived signatures is nontrivial and requires a detailed understanding of the object reconstruction in the individual experiments.  Motivated by these challenges, we outline a strategy for the systematic study of various long-lived neutral particle collider signatures that is based on simplified models, without the need to revert to the more complex models that predict such topologies or to recasting canonical $\MET$ analyses.  

The use of simplified models has become one of the main vehicles to characterise collider searches for new particle production at the LHC.
Today, the majority of searches for SUSY and DM production in collision data are benchmarked using this bottom-up approach. In this work we will focus on the recently established simplified models for DM searches~\cite{Boveia:2016mrp, Abercrombie:2015wmb}, which are based on previous work~\cite{Petriello:2008pu,Gershtein:2008bf,Dudas:2009uq,Bai:2010hh,Fox:2011pm,Goodman:2011jq,An:2012va,Frandsen:2012rk,Dreiner:2013vla,Cotta:2013jna,Buchmueller:2013dya,Abdullah:2014lla,Buchmueller:2014yoa,Abdallah:2015ter,Malik:2014ggr,Buckley:2014fba,Harris:2014hga,Buchmueller:2015eea,Haisch:2015ioa,Abdallah:2015ter,Fox:2012ee}.
In our approach, we extend these models to include collider signatures of long-lived neutral particles. While the primary vertex of the interaction is described using standard simplified DM model, the secondary displaced vertex arising from the decay of the long-lived particle has its own model, which is parameterised in terms of a handful of basic parameters and interactions.  The decay model could involve light ``decay mediator" particles, or simply an Effective Field Theory (EFT) vertex.  This modular approach facilitates event simulation, which is an important requirement for the application of models in experimental analyses.  Furthermore, using the simplified models for long-lived particle signatures, we show examples of how these models can be motivated by and mapped onto UV-complete models, such as gauge-mediated SUSY breaking and models of neutral naturalness.  Finally, although simplified DM models form the foundation of these extensions for neutral long-lived particles, we will also argue that this strategy can be extended to other physics scenarios (e.g. long-lived charged particles).
 
The paper is structured as follows: In Section~\ref{sec:simpDMintro} and Section~\ref{sec:Displaced} we first outline the extension of simplified DM models to include a variety of long-lived particle signatures. In Section~\ref{sec:exp} we discuss the relevant experimental signatures and provide recommendations on how to apply these simplified models in experimental data analysis, while in Section~\ref{sec:models} we motivate this approach by showing how these models can be mapped onto complete models and theories.  We then discuss possible extensions to our approach to also include other long-lived signatures  in~Section~\ref{sec:extension} and end with our conclusions in Section~\ref{sec:conclusion}.  We also provide a hands-on description of the model implementation in event generators in Appendix~\ref{A} and demonstrate the utility of this modular approach with some sample event distributions generated in this way.
 
{\section{Simplified Model Framework} \label{sec:simplified}
\subsection{Production from Simplified DM Models}
\label{sec:simpDMintro}
The programme of general DM searches at the LHC and Tevatron began with interpretations of the results in the framework of an EFT approach~\cite{Beltran:2010ww,Goodman:2010yf,Bai:2010hh,Goodman:2010ku,Rajaraman:2011wf,Fox:2011pm}  in which all the possible effective operators allowed by symmetry and dimensional analyses are considered.  However, as pointed out by several independent groups~\cite{Bai:2010hh,Fox:2011pm,Goodman:2011jq,Shoemaker:2011vi,Fox:2012ee,Buchmueller:2013dya,Busoni:2013lha,Busoni:2014sya,Busoni:2014haa}, there are some drawbacks to this approach. In particular, by design the EFT approach only considers the SM and dark matter particles as light degrees of freedom, thus if there are additional light particles, such as a light mediator, then an alternative description of the phenomenology is required.  As a result, efforts have shifted towards an alternative simplified model paradigm that includes additional mediators.  This has led to an extensive effort amongst both theorists and experimentalists at the LHC to establish a systematic programme to characterise DM searches using simplified models~\cite{Boveia:2016mrp,Abercrombie:2015wmb}.

The current simplified models used to interpret DM searches at the LHC typically contain four independent parameters: the mass of the DM particle, $m_\chi$,
the mass of the mediator, $m_\phi$, the coupling of the mediator to the DM particles, $g_\chi$, and the coupling of the mediator to quarks, $g_\phi$.  A minimal width is also often assumed. For the latter, as a simplifying assumption, the mediator is assumed to couple to all quark flavours with equal strength. The DM particle ($\chi$) is assumed to be either a Dirac fermion, Majorana fermion, or a real or complex Scalar. The mediator exchange can either take place in $s-$ or $t-$channel and the interaction structure of the mediator is chosen to be either a Vector, Axial-Vector, Scalar or Pseudoscalar interaction. Table~\ref{tab:parameters} gives an overview of the different building blocks that form simplified DM models that are currently used to interpret DM searches at colliders (see ~\cite{Boveia:2016mrp,Abercrombie:2015wmb} for further details).
\begin{table}
\centering
\begin{tabular}{ |c||c||c|}
\hline
\multicolumn{3}{ |c| }{Simplified DM Models } \\
\hline \hline
Variables & DM candidate & Interaction \\ \hline
 $m_\phi$ & Dirac & Vector \\ 
 $m_1$ & Majorana & Axial-Vector \\
 $g_\chi$ & Scalar-real & Scalar \\
 $g_\phi$ & Scalar-complex & Pseudoscalar \\ \hline 
 \multicolumn{3}{ |c| }{Displaced Signature Extension} \\ \hline 
 $\tau$ , $m_2$  &  \multicolumn{2}{ |c| }{ Decay of $\chi_2 \rightarrow \chi_1 X$} \\ \hline
\end{tabular}
\caption{Overview of the different building blocks that form simplified DM models. The lower part of this table lists the kinematic variables, lifetime ($\tau$) and mass ($m_2$) of the excited state $\chi_2$ and its decay $\chi_2 \rightarrow \chi_1 X$, which are required to add the displaced signature to the standard simplified DM models.}
\label{tab:parameters}
\end{table}

The approach advocated here is to employ these same simplified models to capture the phenomenology of long-lived neutral particles.  Referring to these long-lived particles as $\chi_2$, then the central red dot in Fig.~\ref{fig:disp} is a schematic for the $\chi_2$ production encoded by the dark matter simplified models.  Importantly for triggering considerations, this also economically enables the inclusion of initial state radiation in event generation in the same way as for the mono-X signatures required of DM searches.  In this approach, the simplified models for long-lived neutral particles will share the same strengths and shortcomings as when applied to dark matter production.

\begin{figure*}[]
  \centering
  \includegraphics[width=0.7\textwidth]{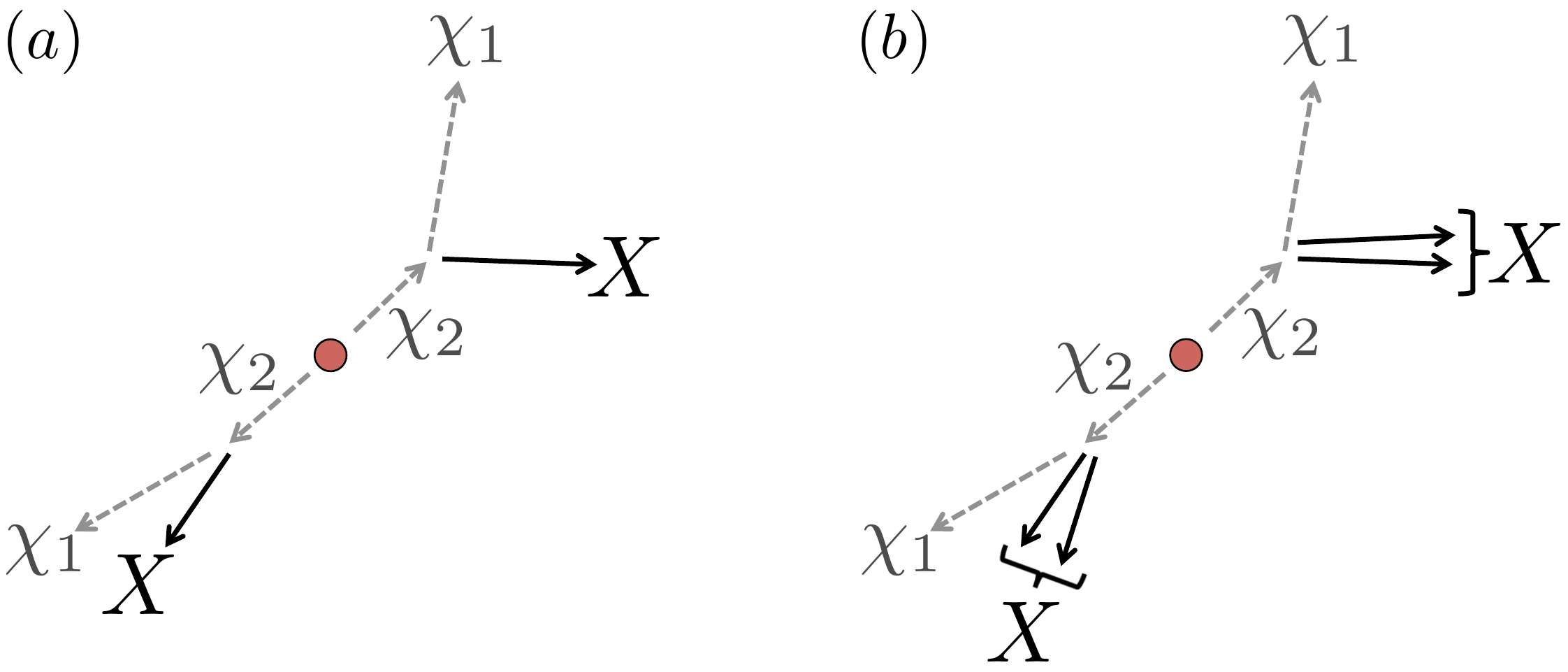}
  \caption{Collider signatures of displaced DM.  (a)  A pair of displaced vertices is observed with a single state $X$ produced at each vertex.  (b)  A pair of displaced vertices is observed with a collection of states $X$ produced at each vertex.  In both cases the DM will typically carry away missing transverse energy, which is a smoking gun for displaced DM production.}
  \label{fig:disp}
\end{figure*}

On this latter point, although the simplified model approach has proven to be very useful to perform a systematic and characterisation of DM searches at colliders, there are well-known limitations to this framework that will also apply to the displaced vertices framework studied here. In particular, comparing simplified models with more UV-complete models, if there are any signatures present in the complete model as a result of the richer spectrum of states, such as cascade decays in SUSY scenarios, then the simplified models strategy may miss signatures that turn out to be the most constraining.  As a result we would advocate the simplified models approach as complementary to the study of complete models, but not as a comprehensive substitute.
%
\subsection{Displaced Decay Models} \label{sec:Displaced}

While the DM simplified models economically describe the production of the long-lived state $\chi_2$, we must also describe its decays.  
To this end we must add at least one new degree of freedom, although it may be desirable to add more.  
As noted, the usual DM candidate in simplified models, $\chi$, is now identified as an excited dark sector state, $\chi_2$. Therefore, we must add to the model the true stable DM state $\chi_1$.
We will now describe how the decays of $\chi_2$ may be modeled through explicit simplified decay models or through an EFT framework, starting with the latter.
\subsubsection{Decay EFT}
\label{sec:decayEFT}
By construction, any effective theory includes the degrees of freedom considered relevant below some cutoff scale, $\Lambda$, and allows for all local operators consistent with the symmetries of the theory.  Often a single class of operators is considered in a specific process for economy.  This effective theory will break down at scattering energies $E \gtrsim \Lambda$, which manifests itself through unitarity violation in scattering amplitudes or, equivalently, by arbitrary operators of a larger dimension giving a comparable contribution at this energy scale.

Let us consider the scenario where $\chi_1$ is the only additional degree of freedom which is added to the DM simplified model. We must introduce an additional coupling, $g_{12}$, to $\chi_2$ and SM states to allow the decay $\chi_2 \to \chi_1 X$. Here, $X$ is a SM object, which could either be individual or multiple particles.  This setup is depicted in Figure~\ref{fig:disp}.  We have now introduced two additional parameters: the mass of the DM state $m_1$ and the EFT scale $\Lambda$, which dresses the specific interaction enabling the decay of $\chi_2$, and in which the coupling $g_{12}$ may be absorbed.  However, in terms of model inputs $\Lambda$ can be traded instead for the more physical parameter $\tau$, which describes the lifetime of the excited state, or its width $\Gamma$, as follows. 

The lifetime of the excited state $\chi_2$ depends on the masses of the particles involved and their couplings, and is given by
\be
\tau^{-1} = \Gamma =  \frac{1}{2 m_2} \int d\text{LIPS}_f \left| \mathcal{M} (m_2 \to \{p_f \} ) \right|^2 
\label{eq:lifetime}
\ee
where $\mathcal{M}$ is the matrix element, proportional to $g_{12}$ and $d\text{LIPS}_f$ is the Lorentz invariant phase space. 
We can parameterise the operator, and therefore the matrix element as $\Lambda^{-d} {\cal O}_{12}$ where $(4+d)$ is the dimension of the operator. Such an operator schematically gives us 
\beq
\Gamma \sim \frac{(m_2-m_1)^a}{\Lambda^{2d}}m_2^{2d +1-a},
\label{eq:eff2body}
\eeq
where the exponent $a$ depends on the specific form of the operator ${\cal O}_{12}$. 

The EFT description of the displaced decays described above is typically valid for the following reasons. We will always consider mass splittings, and hence visible energy deposits, from the displaced decays that are large enough to give an observable signature in LHC detectors, $m_2-m_1 \gtrsim \mathcal{O} (\text{10s GeV})$ and particle masses accessible at the LHC $m_2 \lesssim \text{TeVs}$.  By comparing these two energy scales, one sees that a displaced decay requires $\Gamma \ll m_2$ which, through Eq.~\ref{eq:eff2body}, implies that $\Lambda \gg m_2-m_1$.  Since the energy flowing through the effective operator in this decay is $\mathcal{O}(m_2-m_1)$, the matrix element will never approach $|\mathcal{M}| \sim 1$. Therefore, it will always be a good approximation to employ only the lowest dimension operator.  In addition, higher orders in $\left[ (m_2-m_1)/\Lambda\right]^n$ will always be smaller than the leading order pieces.

Since we are considering the scenario in which both $\chi_1$ and $\chi_2$ are completely SM gauge neutral, the relevant operators are similar to those considered previously for DM at colliders.  The displaced vertices will thus be characterised by additional operators, $\chi_2\chi_1\times\cal{O}_{\rm SM}$ ($\phi_2\phi_1\times\cal{O}_{\rm SM}$), for fermionic (scalar) DM, respectively.  A complete treatment of all effective operators coupling pairs of neutral states to the SM would require a detailed treatment that, for example, ensures that a complete basis of operators is considered.  Alternatively some preferred choice of operator basis could be motivated by considerations related to the theory in the UV.  Both of these options are beyond the intended scope of this work. Instead, we will characterise the decays by the SM object in the final state. We provide a sample of operators in 
Table~\ref{table:displacedoperators}. 
\begin{table}[h]
\begin{center}
\begin{tabular}{|c|c|c|}
\hline
final state X &$\mathcal{O}_F$&$\mathcal{O}_S$\\
\hline
$\gamma$ &$\frac{1}{\Lambda}\overline{\chi}_2 \sigma_{\mu\nu} \chi_1 F^{\mu\nu}$ & $\frac{1}{\Lambda^2}(\phi_2 \partial_\mu \partial_\nu \phi_1) F^{\mu\nu}$ \\
$Z$ &$\frac{1}{\Lambda}\overline{\chi}_2 \sigma_{\mu\nu} \chi_1 Z^{\mu\nu}$ &$\frac{1}{\Lambda^2}(\phi_2 \partial_\mu \partial_\nu \phi_1) Z^{\mu\nu}$\\
$h$ &$\overline{\chi}_2 \chi_1 h$ &  $\Lambda \phi_2 \phi_1 h$ \\
$jj$  &$\frac{1}{\Lambda^3}\overline{\chi}_2 \chi_1 \Tr[G^{\mu\nu} G_{\mu\nu} ]$ & $\frac{1}{\Lambda^2}\phi_2 \phi_1 \Tr[G^{\mu\nu} G_{\mu\nu} ]$ \\
$\overline{l} l$  &$\frac{1}{\Lambda^2}\overline{l} l \overline{\chi}_2 \chi_1$ & $\frac{1}{\Lambda}\phi_2 \phi_1 \overline{l} l$ \\
$\overline{b} b$  &$\frac{1}{\Lambda^2}\overline{b} b \overline{\chi}_2 \chi_1$ & $\frac{1}{\Lambda}\phi_2 \phi_1 \overline{b} b$ \\
$\overline{t} t$  &$\frac{1}{\Lambda^2}\overline{t} t \overline{\chi}_2 \chi_1$ & $\frac{1}{\Lambda}\phi_2 \phi_1 \overline{t} t$ \\
 \hline
\end{tabular}
\caption{List of example effective operators for the decay $\chi_2 \rightarrow \chi_1 X$ for fermionic (middle column) and scalar (right column) DM particles. Each of these operators corresponds to different final state X (left column). Note that this is not an exhaustive list. For example, one could also have diboson final states.} 
\label{table:displacedoperators}
\end{center}
\end{table}%

\subsubsection{Decay Simplified Models}
\label{sec:decaymed}
The previous discussion on EFT considered the case where $\chi_1$ was the only new light degree of freedom. If in addition to $\chi_1$ there is a new light field below the cutoff $\Lambda$ that participates in the $\chi_2$ decays then it must also be added to the theory.  Let us call this field the ``decay mediator" $\phi_D$.  
The addition of models with light decay mediators is phenomenologically motivated as they may have distinctive signatures.

Let us consider a light decay mediator $\phi_D$, of mass $m_D \ll m_2-m_1$.  As depicted in Fig.~\ref{fig:lightmediator}, in this instance the decay process will be two-step, $\chi_2 \to \chi_1+\phi_D (\to SM+SM)$, where $\phi_D$ decays promptly.  $\phi_D$ can be boosted, leading to a new class of displaced objects: boosted pairs of SM fields.  For example, if $\phi_D$ decayed to boosted  muon pairs then each displaced vertex would feature highly collimated muons. 

\begin{figure}[htbp]
\begin{center}
\includegraphics[width=0.3\textwidth]{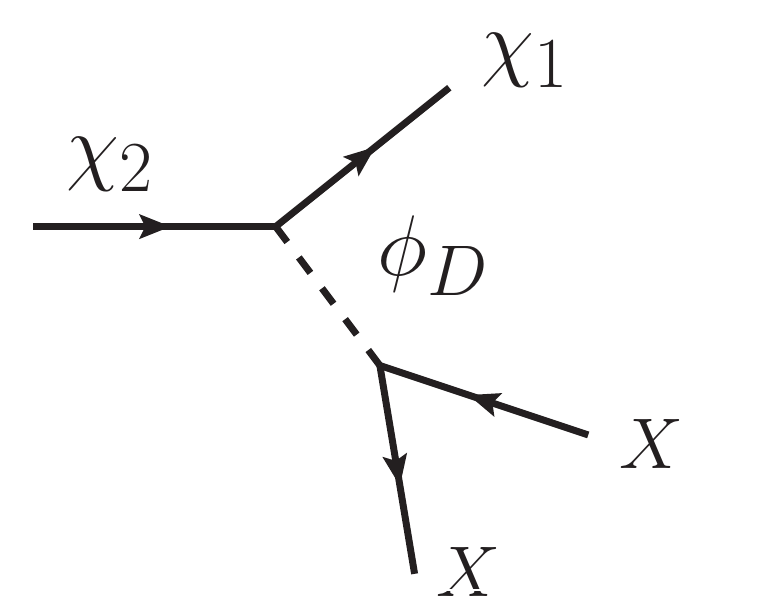}
\caption{Topology for the decay of $\chi_2$ into $\chi_1$ and SM particles ($X$) through a light mediator $\phi_D$.}
\label{fig:lightmediator}
\end{center}
\end{figure}

To construct models of decay mediators one may again take inspiration from the DM simplified models.  For EFT models of decays we could take the DM EFT vertices and make the replacement $\chi^2 \to \chi_1 \chi_2$.  Similarly, for decay simplified models, we may take the DM simplified models coupling DM pairs to a mediator $\phi$ and make the replacement $\chi^2 \to \chi_1 \chi_2$ and $\phi \to \phi_D$, including the mediator interactions with the SM fields.

\begin{table}[h]
\begin{center}
\begin{tabular}{|c|c|c|}
\hline
final state &$\mathcal{O}_{DM} + \mathcal{O}_{SM}$\\
\hline
$$  &$ -g_{12} \phi_D^{\mu} \overline{\chi}_1 \gamma_\mu \chi_2 - g_q \phi_D^{\mu} \overline{q} \gamma_\mu q$ \\
$$  &$ -g_{12} \phi_D^{\mu} \overline{\chi}_1 \gamma_\mu \gamma_5 \chi_2 - g_q \phi_D^{\mu} \overline{q} \gamma_\mu \gamma_5 q$ \\
$\overline{f}f$  &$ -g_{12} \phi_D \overline{\chi}_1 \chi_2 - g_q \phi_D \overline{q}  q$ \\
$$  &$ -i g_{12} \phi_D \overline{\chi}_1 \gamma^5 \chi_2 - g_q \phi_D \overline{q} \gamma^5 q$ \\
 \hline

\end{tabular}
\caption{A small sample list of example vector, axial-vector, scalar, and pseudo-scalar decay mediator couplings for fermionic DM particles.  Similar models may also be constructed for bosons.} 
 \label{table:displacedmediators}
\end{center}
\end{table}%
Following the models discussed in Ref.~\cite{Boveia:2016mrp,Abercrombie:2015wmb}, we present a list of possible decay mediator models in Table~\ref{table:displacedmediators}. 
 Note that these decay mediator models have no limit that captures the mono-boson decays of the first three EFT operators in Table~\ref{table:displacedoperators}, 
and the EFT operators, by construction, have no limit that captures the phenomenology of light decay mediators.  Thus, together both classes of models encompass a complementary set of phenomenological possibilities.

\section{Experimental Signatures and Benchmark Scenarios}
\label{sec:exp}

\subsection{Characterisation of long-lived plus $\MET$ signatures using simplified models}
The simplified models proposed in Section~\ref{sec:simplified} give rise to smoking gun collider signatures that directly connect with dark sector physics.  They also establish a systematic search programme by combining the signatures into a set of reference analyses. In the event of the discovery of events exhibiting displaced vertices, these simplified models can also be used to characterise the discovery using different low-level hypotheses, which in turn may suggest new searches required to reveal the underlying nature of the discovery.  The relevant signatures include:
\begin{itemize}
\item $\MET$:  This is typical in DM searches and reveals the escape of neutral DM candidates from the detector.
\item \emph{Displaced vertices}:  By design these models lead to the appearance of displaced vertices in colliders.  The observed objects may be any SM final state that is gauge neutral.  Furthermore, the observed objects may be highly boosted and collimated.
\item \emph{Displaced vertices are paired}:  This connects to the underlying $\mathcal{Z}_2$ symmetry that enforces pair production of dark sector states.  The requirement of coincident pairs significantly reduces backgrounds.\footnote{Events with a single displaced vertex are possible through $\chi_1 \chi_2$ production.  However, this production rate cannot be too large, otherwise the decays, which proceed through the same operator, would not be displaced.}
\item \emph{Non-pointing events/large impact parameter}:  This is an important characteristic feature, used to reveal additional evidence for DM in the final state.  Even in the event of vanishing MET the visible final states in each displaced vertex typically will not point back towards the interaction vertex or beam pipe, providing a complementary source of evidence for DM.
\item \emph{Initial state radiation}:  In the event that the visible decay products are too soft for triggering, mono-X ISR signatures typical in DM searches may also be used, in tandem with the other signatures, for event characterisation and determination of dark sector couplings.
\end{itemize}
All of these signatures provide useful handles to search for displaced topologies arising at the DM frontier. The challenge is to combine these individual signatures into a set of analyses that can cover most of the relevant topologies. Since the existing experimental programme contains only a few long-lived searches requiring significant $\MET$, it is reasonable to start with a rather small set of new displaced plus $\MET$ searches and to extend the comprehensive prompt $\MET$ searches to also cover long-lived signatures. 

\subsection{Minimal set of long lived plus $\MET$ searches}\label{DM}

We propose a minimal set of displaced $\MET$ searches (dMETs) that cover the basic displaced SM particles -- quarks, leptons and photons -- in combination with significant $\MET$. Following the logic of Table~\ref{table:displacedoperators}, the neutral displaced $X$ system could consist of any SM particles, including $Z, h, \gamma$ and $f\bar{f}$. However, in the interest of keeping the initial set of benchmarks manageable, we focus the reference analyses on signatures involving displaced pairs of jets and leptons as well as displaced $\gamma$'s. These final states would cover most of the relevant neutral displaced signatures and can be extended to add more dedicated requirements like $Z$ and $h$ kinematic compatibility or heavy quark tagging. Table~\ref{table:dMETs_DM} shows a list of the basic dMETs that are categorised by the SM particles that define the $X$ system in the decay of $\chi_2 \to \chi_1 + X$ as well as by the nature of displaced vertices, which can either be a single or a pair of displaced SM particles. Furthermore, to facilitate the trigger acceptance for these displaced topology searches, especially for soft $X$ systems, the dMETs can be combined with a ISR signature such as an additional hard jet or additional hard $\gamma$. 

\begin{table}[h]
\begin{center}
\begin{tabular}{|c||c|c|c|c|c|c|}
\hline
 \multicolumn{6}{ |c| }{$\MET$ plus displaced $X$ system} \\
\hline 
 dMETs&dMET$_{jj}$ & dMET$_{e^+e^-}$ & dMET$_{\mu^+ \mu^-}$ &  dMET$_{\tau^+ \tau^-}$ & dMET$_{\gamma}$  \\ \hline
 $X$& $jet$-pair & $e$-pair  & $\mu$-pair &  $\tau$-pair & $\gamma$  \\ \hline 
\end{tabular}
\caption{Minimal set of dMETs searches for neutral displaced SM particles. To facilitate the trigger acceptance for these topologies, especially for soft $X$ systems, the dMETs can be combined with an ISR signature, such as an additional hard jet or hard $\gamma$. A list of basic operators that would give rise to such topologies is shown in Table~\ref{table:displacedoperators}.  \label{table:dMETs_DM} }
\end{center}
\end{table}%

While this minimal set of dMETs does not cover \emph{all} potential long-lived particle signatures that can arise in dark sector physics, establishing these analyses would already go significantly beyond the current experimental programme for long-lived plus $\MET$ topologies. At present, the only search that is consistently performed by the experiments is the GMSB inspired dMET$_{\gamma}$. All other displaced final states are either not combined with a $\MET$ requirement or have to be recasted from prompt $\MET$ searches, which in both cases is often not optimal. 

The minimal set of dMETs also demonstrates that using simplified models as a vehicle to motivate, develop, and benchmark new long-lived searches provides a straightforward approach to explore opportunities in the long-lived particle sector more systematically and can serve as the basic signature hypothesis to develop and benchmark the corresponding search (see Section~\ref{sec:extension} for some examples).         


\subsection{Hands-on simulation of long-lived plus $\MET$ simplified models} 
\label{sec:simulation}
In this section, we demonstrate how traditional simplified models can be extended to include displaced signatures in practice.  Here, we focus on the s-channel simplified models of dark matter production that were developed in the LHC Dark Matter Forum~\cite{Abercrombie:2015wmb}.  These models are publicly available through the {\sc FeynRules}-based {\sc DMsimp} package~\cite{Alloul:2013bka,Mattelaer:2015haa,Backovic:2015soa,DMsimp}.  Using these models, we construct a concrete simplified decay model of the type discussed in section~\ref{sec:decaymed}.  A complementary example of a decay EFT ({\it cf:} Section~\ref{sec:decayEFT}) for DM production in a GMSB-like model is considered in Appendix~\ref{A}.  The appendix also contains a more general outline of the proposed simulation strategy.  The techniques we use to introduce displaced decays in DM simplified models can also be extended to simplified SUSY and other BSM models.

We begin by considering the class of spin-1 {\sc DMsimp} models that describe the production of Dirac fermion pairs, $\chi\bar{\chi}$, via a vector or axial-vector mediator, $Y_{1}^{V,A}$, where the superscript denotes the type of interaction.  In the {\sc DMsimp} model the $\chi$ represent stable dark matter particles.  The interactions of the $Y_{1}$ mediator with the SM and DM sectors are given as: 
\begin{align}
\mathcal{L}^{Y_{1}}_{SM} &= \sum_{i,j} [ \bar{d}_{i}\gamma_{\mu}(g^{V}_{d_{ij}} + g^{A}_{d_{ij}}\gamma_{5})d_{j} + \bar{u}_{i}\gamma_{\mu}(g^{V}_{u_{ij}} + g^{A}_{u_{ij}}\gamma_{5})u_{j}  ] Y^{\mu}_{1} \,,\\
\mathcal{L}^{Y_{1}}_{DM} &= \bar{\chi}\gamma_{\mu}(g^{V}_{\chi} + g^{A}_{\chi}\gamma_{5})\chi Y^{\mu}_{1}\,.
\end{align}
\noindent To accommodate displaced decays, we introduce a new stable fermion, $\chi_{1}$, and a new scalar particle, $Y_{0}$, into which the $\chi$ can decay.  The required particles and interactions are in fact already contained in a second class of spin-0 {\sc DMsimp} models. These models describe the interactions of scalar (S) and pseudoscalar (P) mediators with the SM and DM sectors: 
\begin{align}
\mathcal{L}^{Y_{0}}_{SM} &= \sum_{i,j} [ \bar{d}_{i}\frac{y^{d}_{i,j}}{\sqrt{2}}(g^{S}_{d_{ij}} + ig^{P}_{d_{ij}}\gamma_{5})d_{j} + \bar{u}_{i}\frac{y^{u}_{i,j}}{\sqrt{2}}(g^{S}_{u_{ij}} + ig^{P}_{u_{ij}}\gamma_{5})u_{j} ] Y_{0}\,,\\
\mathcal{L}^{Y_{0}}_{DM} &= \bar{\chi}(g^{S}_{\chi} + ig^{P}_{\chi}\gamma_{5})\chi Y_{0}\label{eqn:Y0DM}\,.
\end{align}
\noindent Our new displaced plus $\MET$ model is implemented as the union of the two {\sc DMsimp} models. 
Following the notation of Section~\ref{sec:decaymed}, we redefine the $\chi$ in the spin-1 model as $\chi_{2}$, add the $Y_{0}$ and $\chi$ from the spin-0 model (redefining $\chi$ as $\chi_{1}$), and introduce $\chi_{2} \to \chi_{1}Y_{0}$ decays by replacing Eqn.~\ref{eqn:Y0DM} with: 
%
\begin{align}
\mathcal{L'}^{Y_{0}}_{DM} &=  \bar{\chi_{2}}(g^{S}_{\chi} + ig^{P}_{\chi}\gamma_{5})\chi_{1}  Y_{0} + h.c.\ \,.
\end{align}
\noindent Figure~\ref{fig:Zpfeynman} shows a representative diagram from our model that manifests both $\MET$ from $\chi_{1}$ production and displaced vertices from the decay of $Y_{0}$ into SM particles. The {\sc FeynRules} and a corresponding {\sc UFO}~\cite{Degrande20121201} for this model (henceforth, \texttt{DisplacedDM}) are provided at Ref.~\cite{DisplacedDMgit}.  

\begin{figure}[htbp]
  \centering
\includegraphics[width=0.3\textwidth]{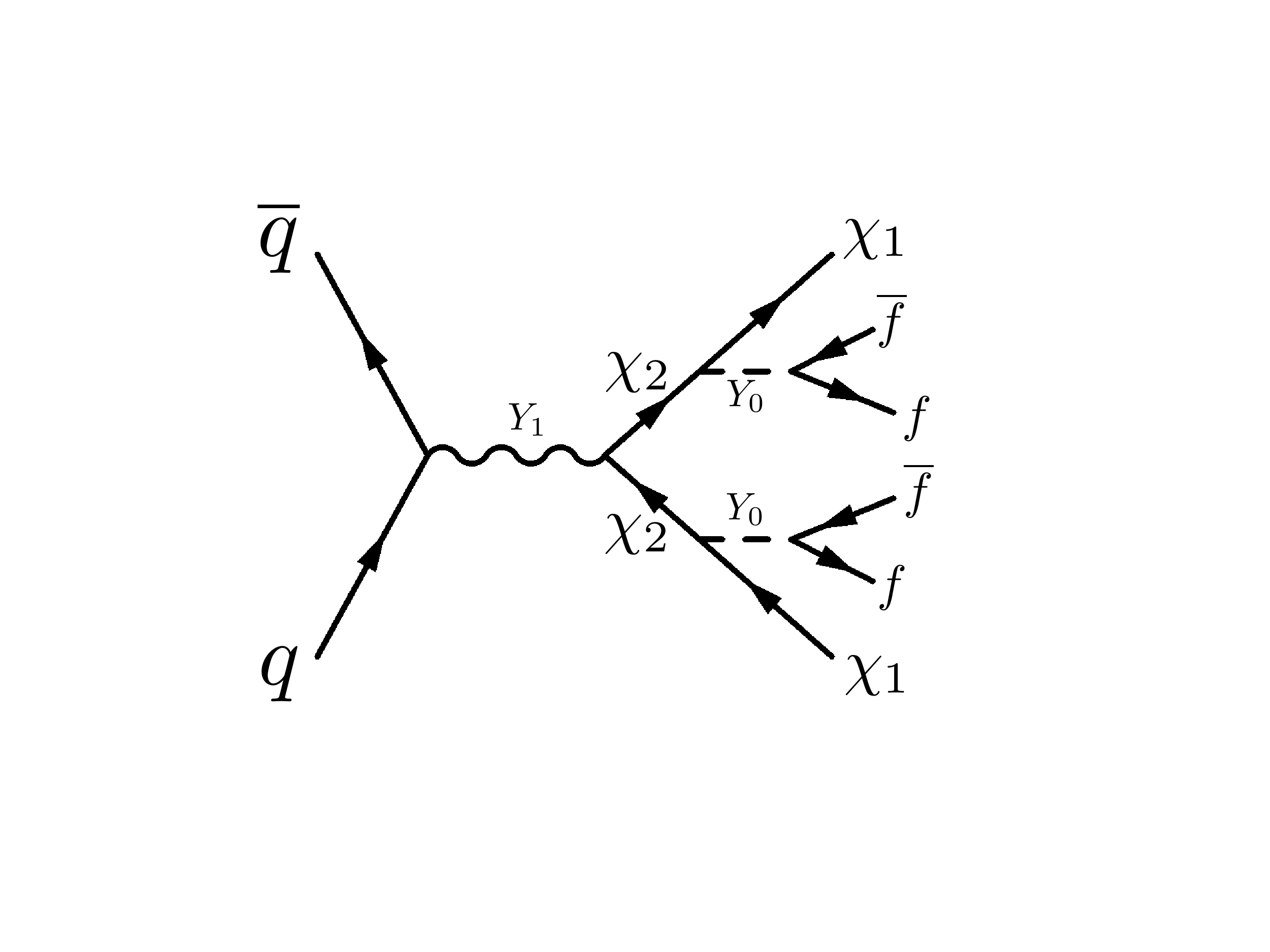}
  \caption{A representative diagram from the \texttt{DisplacedDM} model that produces displaced verticies plus $\MET$. The subscripts on $Y$ indicate the spin of the mediator.}
  \label{fig:Zpfeynman}
\end{figure}

We use the {\sc UFO} to study the relationship between the predicted kinematics and parameter space of the \texttt{DisplacedDM} model.  Simulation proceeds in two stages.  First, the vector-mediated process $pp \to Y^{V}_{1} \to \chi_{2}\bar{\chi_{2}}$ is produced at $\sqrt{s} = 13\rm~TeV$ using \mgvtwo~\cite{Alwall:2014hca} at leading order with the NNPDF3.0 PDF set~\cite{Ball:2014uwa}. We assume flavor universal vector couplings and scan a range of $g_{\chi}$ coupling values and $Y^{V}_{1}$, $Y^{S}_{0}$, and $\chi$ masses.  The partial widths of the $Y^{V}_{1}$, $Y^{S}_{0}$ and $\chi_{2}$ particles are determined automatically in \mg.  Following this approach, we produce LHE~\cite{Alwall:2006yp} files containing unweighted events for the production process.  The lifetimes of the $\chi_{2}$ particles are included using the \texttt{time\_of\_flight} option in \mg.  

For convenience, the subsequent $\chi_{2} \to \chi_{1}Y^{S}_{0} \to \chi_{1}f\bar{f}$ decays are performed using {\sc Pythia8.205}~\cite{Sjostrand:2007gs}.  The masses and widths of the particles in the model are communicated to {\sc Pythia} via the SLHA~\cite{Skands:2003cj,Allanach:2008qq} section of the LHE header.  Spin correlations are not included thus the $\chi_{2}$ and $Y^{S}_{0}$ are decayed isotropically.  In short, the spin-1 {\sc DMsimp} model has been extended with new particles and interactions, while {\sc Pythia} is used to efficiently perform the associated decays.  Alternatively, the new particle content could be added to {\sc Pythia} and the original simplified production model can remain unchanged.  We demonstrate this alternative approach in Appendix~\ref{A}.

The upper-left panel of Fig.~\ref{fig:dxy} shows the distribution of the transverse impact parameter for $Y^{S}_{0} \to \mu^+ \mu^-$ decays for a range of $g^{S}_{\chi}$ coupling values.  The $Y^{S}_{0}$ decays are displaced due to the long lifetime of the $\chi_{2}$ parents.  The upper-right panel shows distributions of the $p_{\rm T}$ of the $\chi_{1}\bar{\chi_{1}}$ system, which is a good generator-level proxy for the $\MET$ observable.  As in the original {\sc DMsimp} models, the shapes of these distributions depend strongly on the mass of the $Y^{V}_{1}$ mediator.   The bottom panel in Fig.~\ref{fig:dxy} shows the difermion mass peak from the resonant decays a 20 GeV $Y^{S}_{0}$ mediator.

\begin{figure}[t!]
  \centering
  \includegraphics[width=0.48\textwidth]{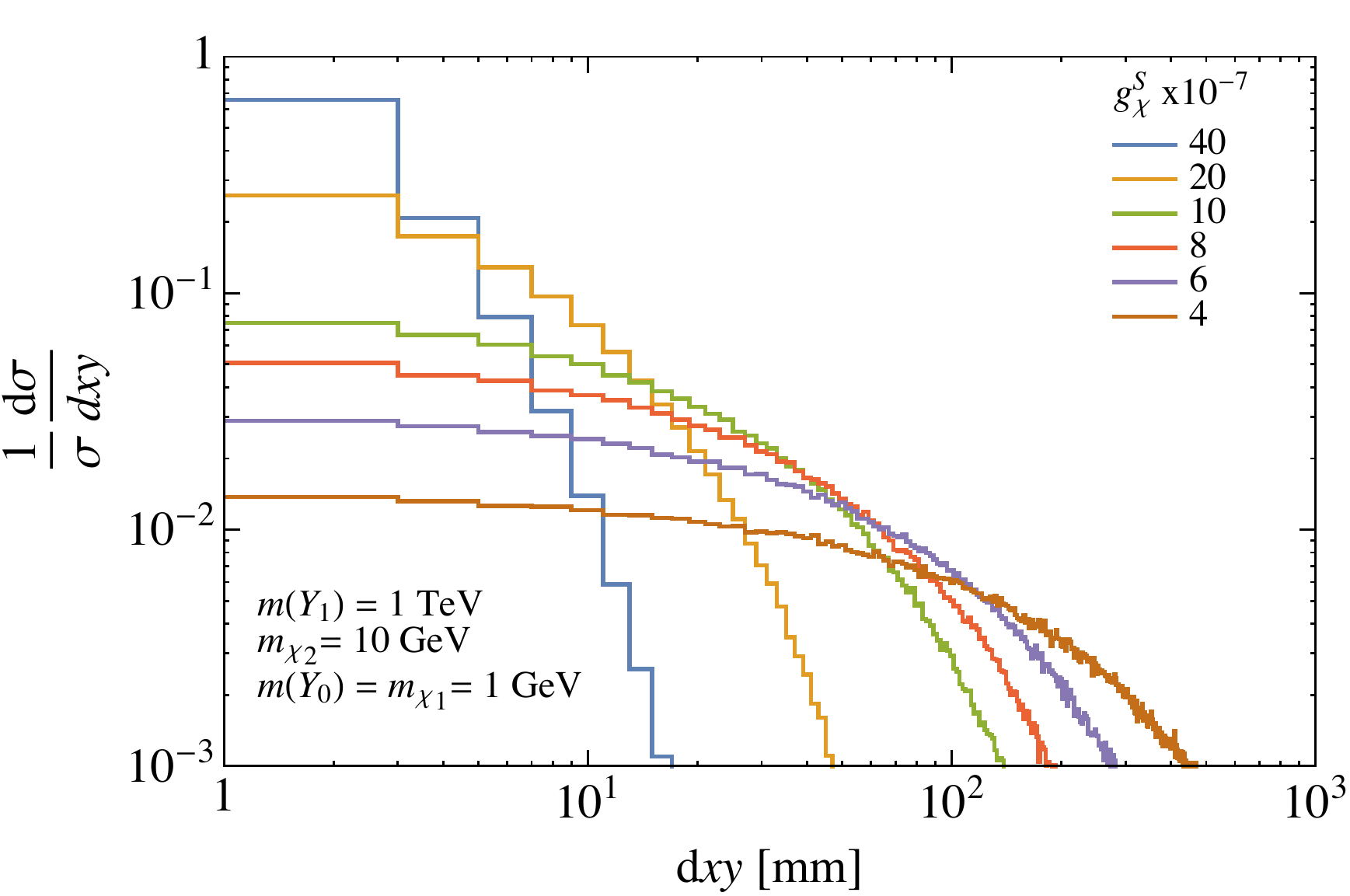}
  \includegraphics[width=0.48\textwidth]{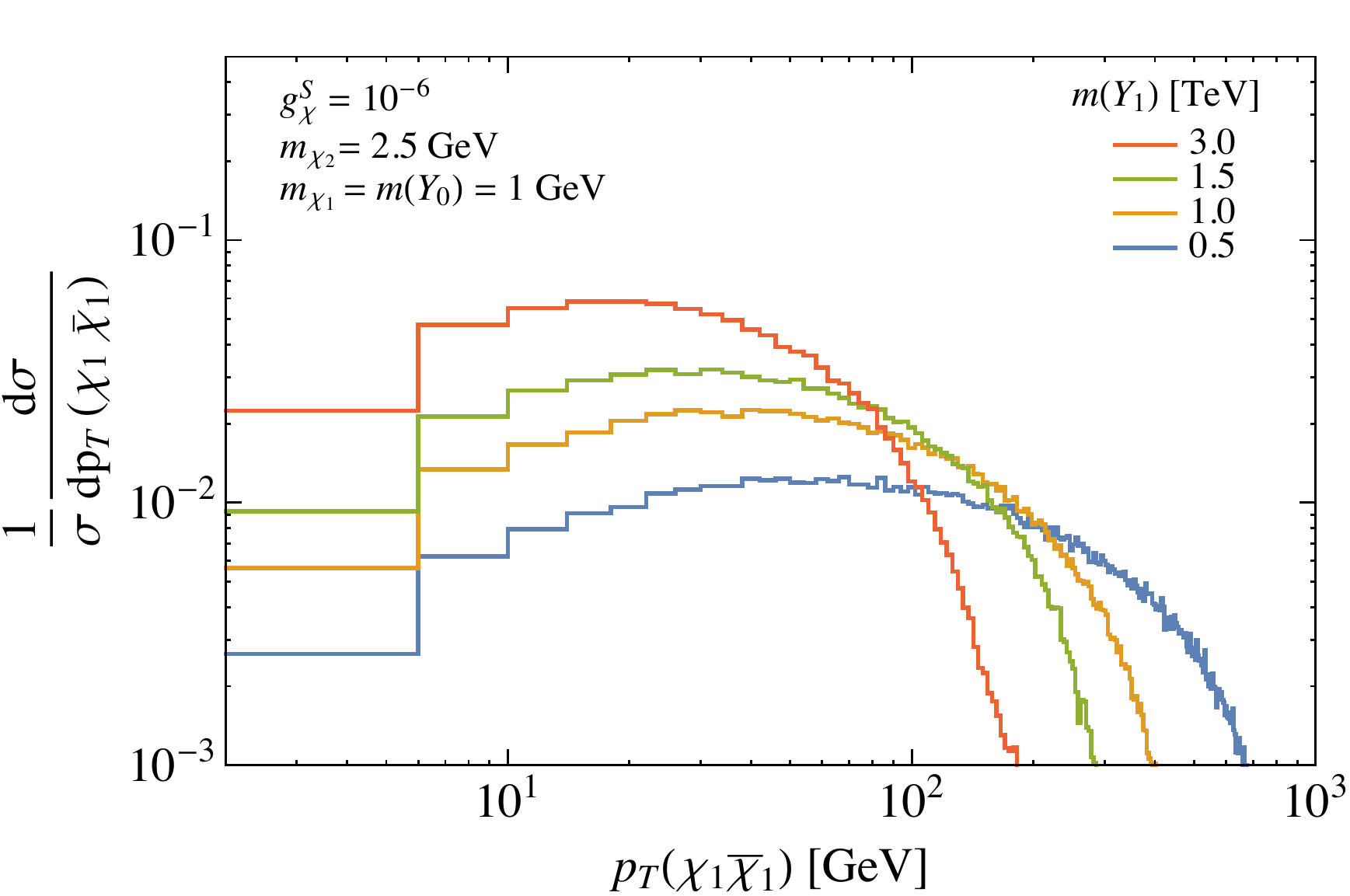}
  ~\\
  \includegraphics[width=0.48\textwidth]{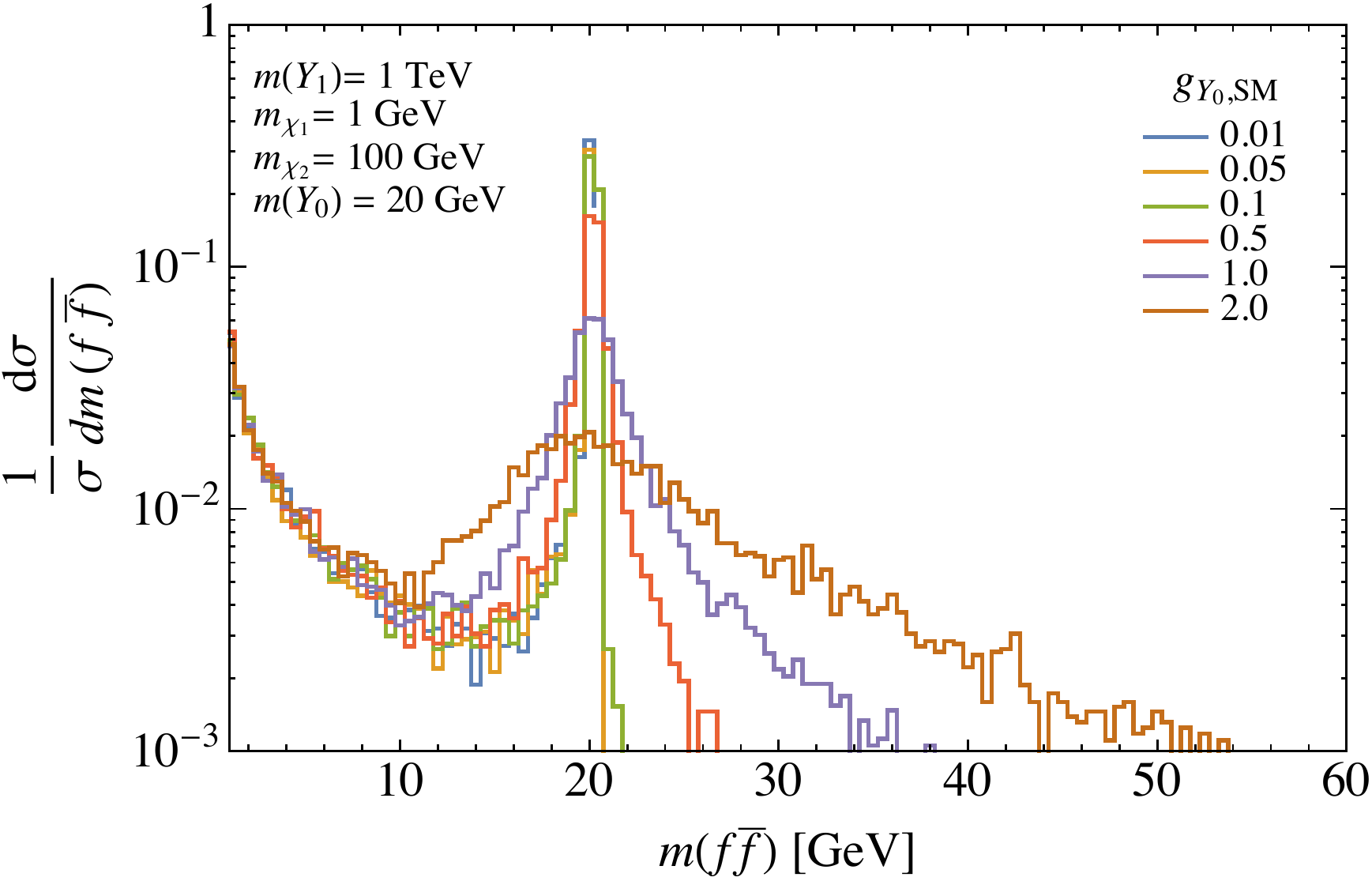}  \caption{Top left: the transverse impact parameter of $Y^{S}_{0} \to \mu\mu$ decays ($d_{xy}$) for a range of $g^{S}_{\chi}$ coupling values.  Top right: the transverse momentum of the DM system ($p_{\rm T}(\chi_{1}\bar{\chi_{1}})$) for various $Y^{V}_{1}$ mediator masses.  Bottom: the dimuon mass spectrum for several values of the $Y^{S}_{0}$-SM coupling. Other parameters in the \texttt{DisplacedDM} model are fixed as per the panel headings.  The distributions in all panels are unit-normalised.}
  \label{fig:dxy}
\end{figure}

Thus the extended simplified models enable the efficient modeling of collider signatures and include the relevant kinematic information necessary for the optimisation and benchmarking of dMETs. In particular, we have illustrated relevant features such as the transverse impact parameter, $\MET$, and kinematic features such as final state resonances for light decay mediators.


\section{Mapping onto UV-complete Models} 
\label{sec:models}
To demonstrate the utility of this displaced vertices framework, we provide two concrete examples that map the simplified model interactions described in the previous section to well motivated models. The first example is for the fermion dipole operator $\overline\chi_2\sigma_{\mu\nu}\chi_1 F^{\mu\nu}$, found in GMSB, while the second example is for the scalar operator $\phi_2\phi_1 h$, which can be found in Higgs portal scenarios such as certain Twin Higgs models.
\subsection{Gauge Mediated Supersymmetry Breaking}
\label{sec:GMSB}
In GMSB, the supersymmetry breaking is mediated by gauge interactions (see~\cite{Giudice:1998bp} for a review) and the LSP is typically the gravitino, that may be very weakly coupled to visible sector particles.  In this section we will demonstrate that GMSB with an approximately pure Bino NLSP and a gravitino LSP maps directly onto one of the simplified models proposed here. In GMSB the mediation of SUSY breaking is via the gauge interactions of chiral messenger superfields, $\Phi$, interacting with the goldstino superfield $X$ as
\beq
W=\lambda_{ij}\bar\Phi_i X \Phi_j.
\eeq
The chiral superfield $X$ acquires a vev along the scalar and auxiliary components,
\beq
\langle X \rangle = M+\theta^2 F
\eeq
where $M$ is the messenger scale while $\sqrt{F}$ is a measure of the amount of SUSY breaking. For most realistic cases, it is appropriate to assume that $F\ll M^2$. An appealing property of GMSB models is that the entire supersymmetric spectrum is determined by the supersymmetry-breaking scale $\Lambda=F/M$, the messenger index $N$, the messenger mass $M$, and $\tan\beta$.

\begin{figure}[tbp]
\begin{center}
\includegraphics[width=0.45\textwidth]{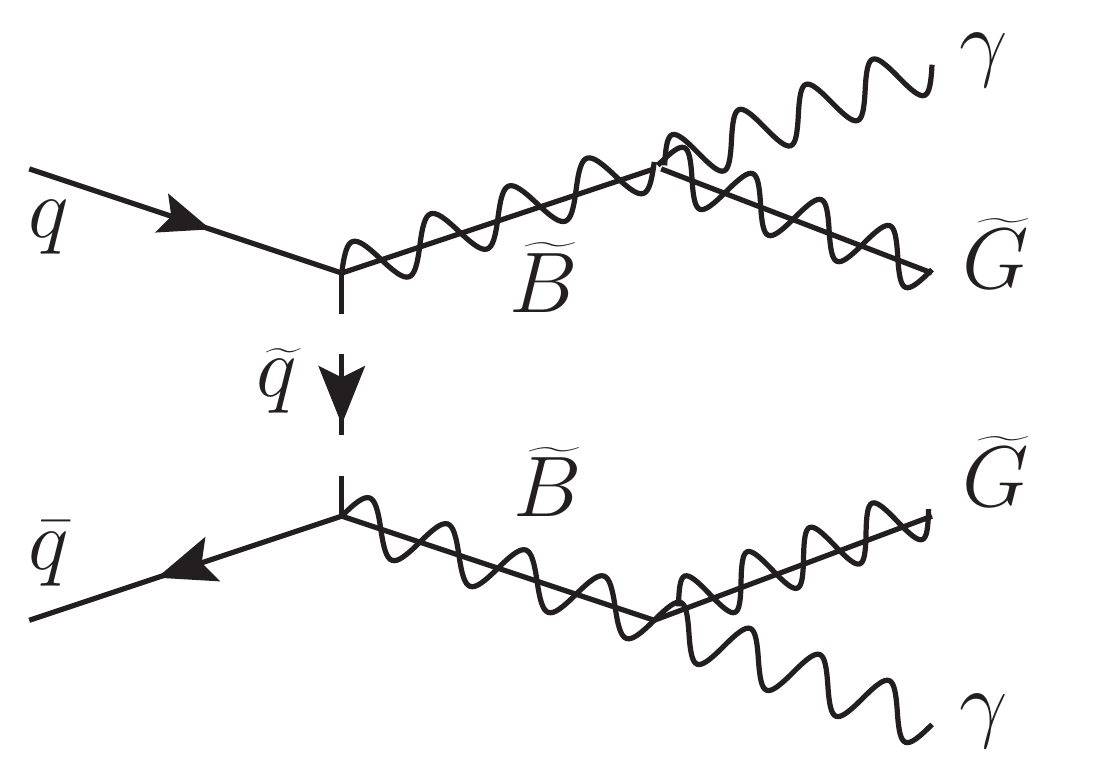}
\caption{Bino production through t-channel squarks, which maps to the t-channel mediators of the simplified DM models.}
\label{fig:GMSB}
\end{center}
\end{figure}

In the leading-log approximation, assuming a single SUSY-breaking superfield satisfying $F\ll M^2$, the supersymmetry breaking Bino mass is
\beq
M_{\widetilde{B}}(t)=\frac{5 \alpha_Y (t)}{12\pi}\frac{F}{M}
\eeq
where we have chosen the number of $SU(5)$ messengers to be $N_5=1$.  The gauge coupling constants, $\alpha_r$, are defined such that $5/3 \alpha_Y\sim \alpha_W\sim \alpha_S$ at the GUT scale. The mass of the gravitino is 
\beq
m_{\tilde G}=\frac{F}{k\sqrt{3} M_p}
\label{eq:mgravitino}
\eeq
where $M_p=(8\pi G_N)^{-1/2}\simeq 2.4\times 10^{18}$ GeV is the reduced Planck mass. 
The constant $k\equiv F/F_0$ is the ratio between the fundamental scale of SUSY breaking, $F_0$, and the scale of SUSY breaking felt by the messenger particles, $F$, and represents the mass splitting inside the supermultiplet. This ratio is traditionally set equal to unity.
Unless the messenger scale approaches the Planck scale $M\to M_P$ the gravitino will typically be the LSP.

In GMSB the interactions of the Bino and the gravitino are described by the effective operator
\beq
{\cal L}=\frac{k}{F} \frac{M_{\widetilde B}}{4\sqrt{2}} \overline{\widetilde{B}} \sigma_{\mu\rho} \widetilde{G} F^{\mu\rho}_Y +h.c.
\eeq
where $\widetilde{G}$ is the longitudinal component of the gravitino and $F^{\mu\rho}_Y$ is the hypercharge field strength. 
Thus the interaction allowing Bino to gravitino decays is described by the first operator listed in Table~\ref{table:displacedoperators}, demonstrating that the effective phenomenology of the decay is captured by the simplified model.

From this effective operator two decay channels arise $\widetilde{B} \to \gamma \widetilde{G}$, and if kinematically accessible, $\widetilde{B} \to Z \widetilde{G}$, with widths given by
\bea
\Gamma(\widetilde{B}\to \gamma + \tilde G) &= &k^2 \frac{\cos^2\theta_W }{16\pi F^2} M_{\widetilde{B}}^5 \label{eq:GMSBwidth}\\
\frac{\Gamma(\widetilde{B}\to Z + \tilde G)}{\Gamma(\widetilde B\to \gamma + \tilde G)} &= & \tan^2 \theta_W \left( 1-\frac{M_Z^2}{M_{\widetilde{B}}^2} \right)^4 ~~.
\eea
The total lifetime thus scales as
\be
\tau \sim \frac{M_{\widetilde B}^3F^2}{(M_{\widetilde B}^2-M_{Z,\gamma}^2)^4}
\ee
where for a fixed value of $M_{\widetilde{B}}$ the value of $F$ can be varied independently, demonstrating that the lifetime and mass may be taken as independent parameters.

The production of a pure Bino neutralino is through $t$-channel squark exchange, which maps directly to the colored scalar $t$-channel model of the DM simplified models paradigm (Fig.~\ref{fig:GMSB}). In summary, the production and decay of a pure Bino NLSP is captured by the set of simplified models proposed here, demonstrating that these simplified models do map onto well-motivated BSM models for displaced vertices. 

\begin{figure}[tbp]
\begin{center}
\includegraphics[width=0.45\textwidth]{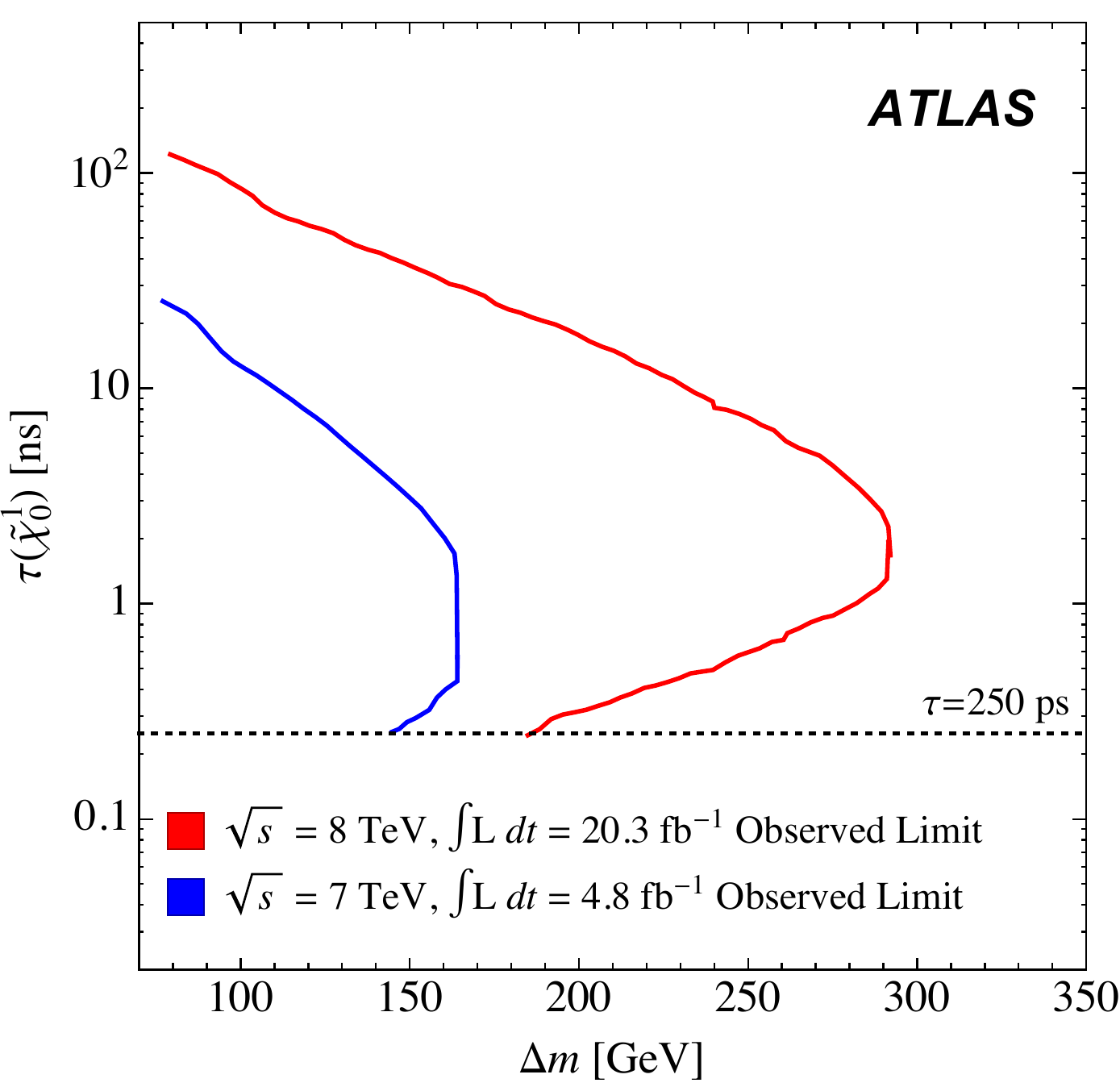}
\caption{Remapping of the ATLAS displaced diphoton + missing energy search into the simplified model variables of Bino lifetime, $\tau$, and Bino-gravitino mass splitting $\Delta m$. The ATLAS search was presented in terms of SPS8 (see text for details).}
\label{fig:ATLAS_GMSB_remap}
\end{center}
\end{figure}

Analyses searching for the displaced decay signature of two displaced photons plus missing energy already exist (e.g.~\cite{Aad:2014gfa,Chatrchyan:2012ir}) and can be converted into the simplified models frame work.  The ATLAS collaboration has performed a search for non-pointing and delayed diphotons and missing transverse momentum ~\cite{Aad:2014gfa}. Here, they present their results in terms of the Snowmass Points and Slopes parameter set 8 (SPS8)~\cite{Allanach:2002nj}, which describes the set of minimal GMSB models in which the NLSP is the lightest neutralino and the LSP is the gravitino, and the other parameters are $N_{mess}=1,~M=2\Lambda, \tan\beta=15$. In this model, the mass difference between $\tilde\chi_0^1$ and $\tilde G$ can be written
\beq
\Delta m\equiv m_{\tilde\chi_0^1}-m_{\tilde G}=\Lambda\left(\frac{5}{12\pi}\alpha_1(t)-\frac{2\Lambda}{\sqrt{3}M_p}\right).
\eeq

Therefore, we can translate the constraints on $\tau(\tilde\chi_0^1)~\rm{vs.}~\Lambda$ into our simplified models parameters $\tau(\tilde\chi_0^1)~{\rm{vs.}}~\Delta m$ (Fig.~\ref{fig:ATLAS_GMSB_remap}).

\begin{table}
\centering
\begin{tabular}{c}
\hline
Map of GMSB to simplified models\\
\hline
$t$-channel, colored scalar mediator\\
$\widetilde B$ Dirac fermion $\to\chi$\\
$m_{\tilde q}\to m_\phi$\\
$m_{\widetilde B}\to m_1$\\
$\sqrt{4\pi\alpha_Y}\to g$\\
$\frac{k}{F} \frac{M_{\widetilde B}}{4\sqrt{2}}\to \tau(\widetilde B)$\\
$\Delta m=m_{\widetilde B}-m_{\tilde G}$\\
\hline
\end{tabular}
\caption{Map of the simplified models parameters to GMSB. The first five rows characterise the DM simplified model while the last two rows are the additional parameters required by the displaced vertices simplified model.}
\label{tab:GMSBparameters}
\end{table}
\subsection{Higgs Portal Hidden Sectors/ Twin Higgs}
Our second example is for the scalar operator $\phi_1\phi_2 h$, which can be found in Higgs portal scenarios. 
The Higgs portal model maps to the $s$-channel scalar mediator DM simplified model, where the scalar mediator is the SM Higgs boson, and in fact would be general enough to cover greater parameter space by considering more general scalar mediators than the Higgs boson alone.
One distinguishing feature of the Higgs portal simplified model for displaced dark matter is that both the $\phi_2$ production and the $\phi_2$ decay are mediated by the same particle.  This is given by
\bea
\mathcal{L} & \supset & \frac{1}{2} \left( \lambda_1 \phi_1^2 + \lambda_2 \phi_2^2 \right) |H|^2 + \lambda_{12} \phi_1 \phi_2 |H|^2 +...\nonumber \\
& \to & \frac{1}{2} \left( \lambda_1 \phi_1^2 + \lambda_2 \phi_2^2 \right) h v + \lambda_{12} \phi_1 \phi_2 hv +...
\label{eq:scalarL}
\eea
where the ellipses denote terms irrelevant for the phenomenology and the second line is in the broken EW vacuum. The couplings $\lambda_1$ and $\lambda_2$ respect independent $\mathcal{Z}_2$ symmetries acting on $\phi_1$ and $\phi_2$. The coupling $\lambda_{12}$ breaks the $\mathcal{Z}_2\times \mathcal{Z}_2$ symmetry to the diagonal, implying that $\lambda_{12}$ can be naturally small.  
\begin{figure*}[t]
\begin{center}
\includegraphics[width=0.8\textwidth]{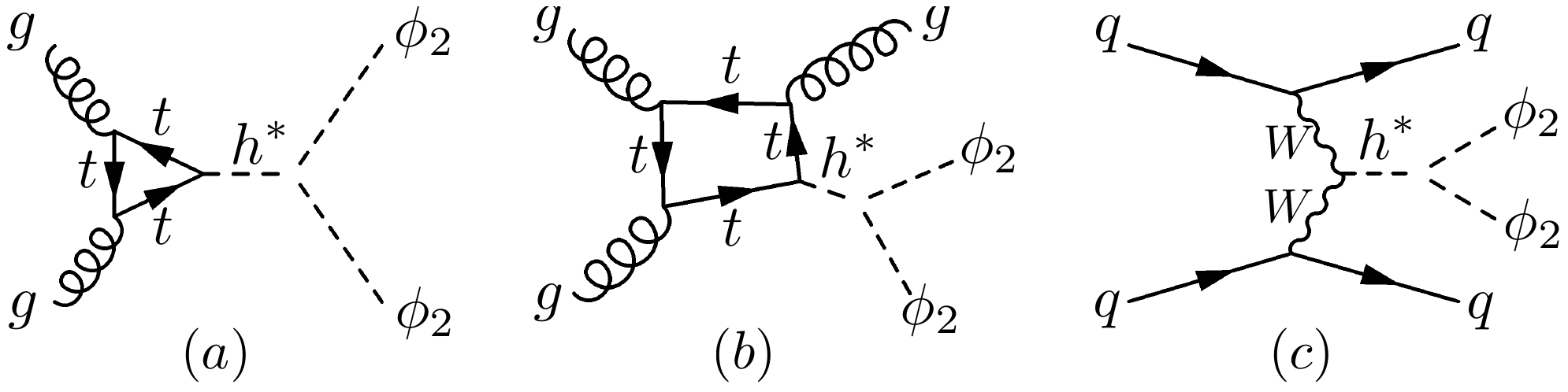}
\caption{Various productions modes for the Higgs-portal modes. Diagrams (a) and (b) show the pair production of $\phi_2$ through the top-loop higgs production, while diagram (c) is through the VBF higgs production.}
\label{fig:hp}
\end{center}
\end{figure*}
The production channel relevant for the Higgs-portal displaced decays is the pair production of the heavier state $\phi_2$ through the Higgs portal (Fig.~\ref{fig:hp}). The associated production of  $\phi_1 \phi_2$ and the pair production of the lighter state $\phi_1$ also proceed through the Higgs portal. However, for $\lambda_{12}\ll1$, the associated production is very suppressed, and $\phi_1$ pair production can only be detected through the standard MET searches, and therefore are not relevant to our purposes. 

\begin{figure}[t]
\begin{center}
\includegraphics[width=0.25\textwidth]{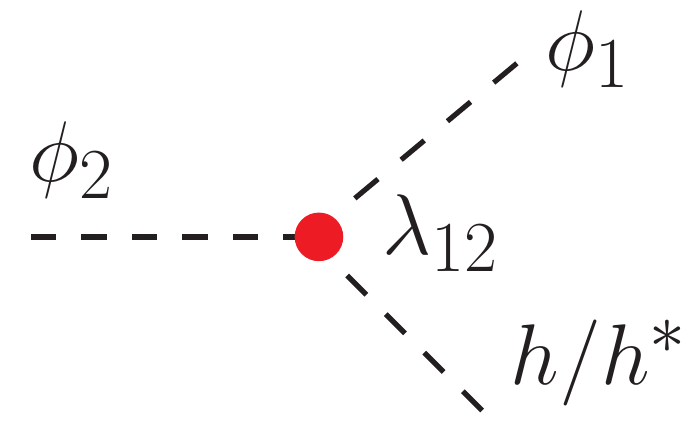}
\caption{Decay of $\phi_2$ through the $\lambda_{12}$ interaction. The emitted Higgs can be on- or off-shell depending on the mass difference of $\phi_2$ and $\phi_1$.}
\label{fig:scalardecay}
\end{center}
\end{figure}

Importantly, $\phi_2$ decays to $\phi_1$ through the Higgs portal alone, emitting an on-shell or off-shell Higgs in the process (Fig.~\ref{fig:scalardecay}).  For small enough $\lambda_{12}$, which is a free parameter, this decay can be displaced.  The on-shell decay width is given by
\be
\Gamma =\frac{\lambda_{12}^2v^2}{16\pi m_2^3}\sqrt{m_h^4+(m_2^2-m_1^2)^2-2m_h^2(m_2^2+m_1^2)}.
\ee
Essentially, in the context of the proposed simplified models, the Higgs boson is the production mediator \emph{and} the decay mediator.  This scenario maps onto the Twin Higgs models~\cite{Chacko:2005pe,Chacko:2005un}, which are a class of ``neutral naturalness" models that have received a recent revival in interest. These models are a class of pseudo-Nambu-Goldstone boson (pNGB) Higgs models that contain two exact copies of the SM, such that the quadratic sensitivity to the cutoff in both sectors is identical.  In addition, there is a scalar potential involving the SM and Twin Higgs doublets
\be
V = \frac{\lambda}{2} \left( |H|^2 + |H^T|^2 - \frac{f^2}{2}  \right)^2  ~~.
\label{eq:portal}
\ee
This scalar potential respects an $\text{SU}(4)$ symmetry, of which the two copies of $\text{SU}(2)_W \times \text{U}(1)_Y$ are subgroups.  As any quadratic corrections are equal for both doublets, they also respect the $\text{SU}(4)$ symmetry.  In the vacuum, this symmetry is spontaneously broken, with $\langle H^T \rangle = f/\sqrt{2}$, and the SM Higgs emerges as a light pNGB.  If the $\text{SU}(4)$ symmetry were exact the SM Higgs would be massless. However, the gauge and Yukawa interactions do not respect the full $\text{SU}(4)$ symmetry, which leads to one-loop quartic scalar potential couplings that give rise to a non-zero Higgs mass.  Importantly, these corrections are only logarithmically sensitive to physics at the cutoff. Thus, these models successfully remove the quadratic sensitivity of the Higgs mass to new physics at the cutoff.

For phenomenology, the most important aspect is that Eq.~\ref{eq:portal} is the only interaction between the SM and Twin sectors,\footnote{If the new physics at the cutoff couples to both sectors then additional couplings are introduced, however this aspect is very model-dependent.} and so tests of the Twin Higgs scenario rely crucially on the collider phenomenology of the Higgs Portal.  In particular, through Eq.~\ref{eq:portal}, when both Higgs bosons obtain vacuum expectation values they will mix, and the SM Higgs inherits couplings to the Twin fields from the Twin Higgs boson.  For example, the neutral component of the SM Higgs doublet is coupled to SM top quarks and Twin top quarks as
\be
\mathcal{L}_{\text{Int}} = \lambda_th \overline{t} t + m_{t_T}\left(1- \frac{h^2}{2 m_{t_T}} \right) \overline{t}_T t_T ~~,
\ee
where the specific form of the interaction arises due to the imposed symmetries.  Notably, just as a loop of top quarks couples the Higgs boson to gluons, similarly a loop of Twin top quarks will couple the SM Higgs to Twin gluons as
\be
\mathcal{L} \supset -\frac{\alpha_3^T}{6 \pi} \frac{v}{f} \frac{h}{f} \text{Tr} G^T_{\mu\nu} G^{T\mu\nu} ~~,
\ee
where $G^T$ is the Twin QCD field strength.  As a result, Twin gluons may be produced in rare Higgs decays.

In the most minimal scenario, the Twin sector only contains the third generation fermions, the so-called ``Fraternal Twin Higgs"\cite{Craig:2015pha}.  In this case, the lightest objects carrying Twin colour are the Twin gluons themselves.  These Twin gluons may hadronise to long-lived glueball states, which may decay back to SM particles back through the Higgs portal~\cite{Juknevich:2009gg,Juknevich:2009ji}.

The lightest Twin glueball state is the scalar $G_{0^{++}}$ (using standard $J^{PC}$ notation), whose mass we denote $m_g$.  This glueball can be pair produced in Higgs decays and subsequently decay back into SM states through the Higgs portal.  If this is the dominant decay mode it can lead to observable rare Higgs decays at the LHC.  However, it is also possible that states such as $G_{0^{++}}$ can decay into further Twin sector states, such as pairs of Twin leptons $G_{0^{++}} \to \overline{\tau}^T \tau^T$.  In this case the decays would be invisible as the $\tau^T$ are SM gauge neutral and would escape the detector unobserved.\footnote{In fact, the Twin leptons are compelling dark matter candidates \cite{Garcia:2015toa,Garcia:2015loa,Farina:2015uea,Craig:2015xla}.}

\begin{table}
\centering
\begin{tabular}{c}
\hline
Map of Fraternal Twin Higgs to simplified models\\
\hline
$s$-channel, scalar mediator\\
$G_{2^{++}}$ scalar $\to\chi$\\
$m_{G_{2^{++}}}=1.4 m_0\to m_\chi$\\
$m_{G_{0^{++}}}=m_0\to m_1$\\
$g_q$ $\to g_\phi$\\
$\frac{\hat\alpha_3 v_h}{6\pi f^2}\to g_\chi$\\
$\frac{y^2\hat\alpha_3}{3\pi M^2}\to\tau(G_{2^{++}})$\\
$\Delta m=0.4 m_0$\\
\hline
\end{tabular}
\caption{Map of twin glueball parameters to the simplified models. The first five rows characterise the DM simplified model while the last two rows are the additional parameters required by the displaced vertices simplified model.}
\label{tab:Twinparameters}
\end{table}

If the lightest glueball decays predominantly into invisible states then it may still be possible to uncover evidence for the hidden sector from the production and decay of excited hidden sector states.  For example, the next lightest glueball $G_{2^{++}}$, with mass $m_2 \sim 1.4 m_0$, is metastable and its dominant decay mode is through radiative Higgs production $G_{2^{++}} \to G_{0^{++}} h$, if the mass splitting permits.  If $m_2 - m_1 < m_h$ then the Higgs is off-shell and the decay will proceed directly to SM fermions or to Twin fermion states.

Depending on the parameters of the model, the $G_{2^{++}}$ can be sufficiently long-lived to give displaced vertices at the LHC.  Rather than going into details on the various transition matrix elements and the details of the lifetime calculation we will refer the interested reader to \cite{Juknevich:2009gg} for relevant expressions.  

It is worth emphasising that although the Twin Higgs provides motivation for the topologies captured by the displaced Higgs portal model, the Higgs-portal model is interesting in its own right due to the simplicity of the setup.  
\section{Avenues for Extensions}
\label{sec:extension}

Throughout we have focused on models where the metastable and stable new states are gauge neutral.  
This was for good reason, as the physics motivation was to search for evidence of dark sectors.  
As we have demonstrated, this scenario allows one to port the well-established dark matter simplified models to models of new, metastable, neutral particles.  
However, models with gauge neutral new states form only a subset of all possible models with displaced vertices. 
Thus, in line with the philosophy of a simplified models programme, one can take our proposed dMETs as the first step to a more comprehensive programme that covers the full range of BSM scenarios.

An appealing possibility for extending this approach is to apply the framework to SUSY searches.
The LHC has already established a strong and diverse analysis programme to search for SUSY particle production accompanied by significant $\MET$. Today, these searches are primarily interpreted in the context of simplified SUSY models~\cite{Barnett:1985ci, Baer:1986ki,Alwall:2008ag, Alwall:2008va}. These simplified models assume a limited set of SUSY particle production and decay modes, but leave open the possibility to vary masses and other parameters freely. Therefore, these simplified models enable comprehensive studies of individual SUSY topologies, and are useful for optimisation of the experimental searches over a wide parameter space without additional limitations on fundamental properties such as masses, production cross sections, and decay modes.    
\begin{figure}[htbp]
\begin{center}
\includegraphics[width=0.65\textwidth]{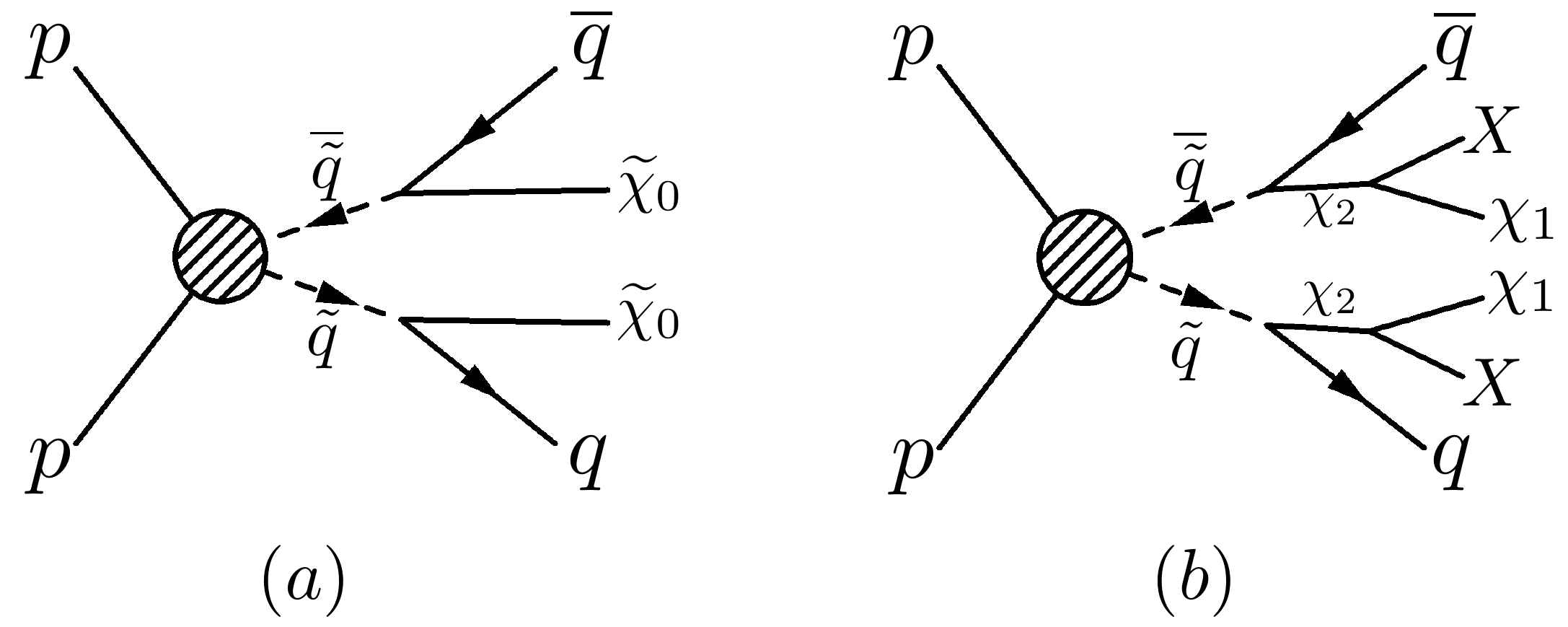}
\caption{ An example of a simplified SUSY model diagram showing the direct production of light flavour squark pairs decaying to two jets plus $\MET$\ ($\squark \squark \rightarrow qq \ninoone \ninoone$) is shown in a). In this framework the decay of  $\squark \rightarrow q \ninoone$ is assumed to be prompt. In b) we show a potential extension of this simplified SUSY model to feature displaced signature. Here the neutralino is replaced with the long-lived fermion $\ninoone \to \chi_2$, which may then decay to SM states $X$ and $\chi_1$.}
\label{fig:SUSY}
\end{center}
\end{figure}

As with the canonical simplified DM models, the vast majority of simplified SUSY models also assume a prompt decay to the weakly interacting stable particle state which in our notation is $\chi_1$.  For example, one simplified model diagram for the direct production of light flavour squark pairs decaying to two jets plus $\MET$\ ($\squark \squark \rightarrow qq \ninoone \ninoone$) is shown in Figure~\ref{fig:SUSY} a).  However, it is well known (see e.g.~\cite{Bauer:2016gys,Bagnaschi:2015eha} for further discussion) that in many SUSY co-annihilation scenarios, the next-to-lightest supersymmetric particle (in our notation $\chi_2$) may have a mass only slightly greater than that of the lightest supersymmetric particle ($\chi_1$), in which case it may have a long lifetime. This opens up the possibility of signatures from displaced vertices in the detector. 

Using the approach outlined in this paper, it would be straightforward to extend the simplified SUSY models to also include long-lived signatures by adding the long-lived particle $\chi_2$ as in the simplified DM models. In the example case provided here, the lightest SUSY state $\ninoone$ would be replaced with a next-to-lightest state $\chi_2$ that can possess a significant lifetime and will decay to $\ninoone$ and $X$, which is shown in Figure~\ref{fig:SUSY} b).  Depending on the choice of SUSY particle for $\chi_2$, the arising final state could either consist of a neutral or charged $X$ system, and may be accompanied with significant $\MET$ (see. e.g.~\cite{Khoze:2017ixx,Mahbubani:2017gjh} for recent work in this direction).  Simulation of the extensions of the usual simplified SUSY models to include long-lived signatures is straightforward and similarly follows the procedure described in Section~\ref{sec:simulation}. Since the corresponding prompt SUSY searches are already in place, adding the long-lived signature to these analyses seems a natural extension of their scope, broadening the long-lived plus $\MET$ search programme at the LHC.    

Furthermore, the simulation approach does not require $\chi_1$ to be a neutral weakly interacting particle. Hence in this vein it would be possible to also establish long-lived simplified models that can benchmark displaced signatures without $\MET$ in the final state, opening the possibility for a systematic characterisation of the long-lived particle frontier in general.     

It should always be kept in mind with any simplified model programme that nature may not abide by our desire for simplicity.  For example, in many SUSY scenarios with metastable particles the dominant signatures could arise from cascade decays involving, for example, heavy gluinos.  Presumably, if the number of simplified models is to remain finite, there will always be examples of BSM models whose phenomenology cannot be captured by simplified models.  Thus the models, and possible extensions, proposed here should be considered as complementary to, but not replacing, model-inspired searches.

\section{Conclusions}
\label{sec:conclusion}
No stone should be left unturned in the search for physics beyond the Standard Model at the LHC.  Some avenues for discovery are well established, however others remain to be fully exploited.  In particular, it is becoming increasingly clear that many opportunities remain to discover long-lived particles from characteristic displaced vertex signatures.  In some cases this will remain true throughout LHC operation, as the distinctive signatures and low backgrounds inherent in displaced vertex signatures will shield the searches from becoming limited by systematic uncertainties.

Further motivation for extending the displaced vertices searches at the LHC comes from the fact that long-lived particles show up readily in a plethora of well-motivated scenarios beyond the Standard Model, from the dark sectors to SUSY to strongly coupled theories to generic hidden valleys. Exploiting this rich frontier of discovery opportunities will require a similarly rich experimental programme.  

However, the experimental long-lived programme today is still rather sparse and does not cover systematically all relevant signatures and final states. To expand this important search programme in the future, simply providing a list of final states is insufficient to craft new experimental searches as signal event simulation plays a critical role in defining, optimising, and interpreting a new search. On the other hand, it is not feasible to generate events for every BSM scenario with displaced vertices.  Thus the theoretical landscape motivates a greater breadth of displaced vertex searches, yet the experimental landscape requires efficient event simulation and modeling. 

To bridge this gap, in this work we have proposed a first set of displaced vertex simplified models connecting these signatures with the dark sector, to cast a wider net in the hunt for dark matter.  We have illustrated the analysis chain from the practical implementation of the simplified models to event generation.  We have also proposed a new set of generic dMET (displaced+MET) searches that can be readily implemented.  Furthermore we have demonstrated how these simplified models and corresponding dMET searches map onto well motivated theories.

The simplified models proposed here do not comprehensively cover the displaced frontier at the LHC, but are rather a first step with a view to connecting with the collider dark matter hunt.  To capture a greater variety of signatures these simplified models will need to be extended and we have outlined how the current set of SUSY simplified models may be used to broaden the scope of the displaced vertices simplified models significantly.

\section*{ACKNOWLEDGMENTS}
We would like to thank James Beacham, David Curtin, Paddy Fox, Kentarou Marwatari, Stephen Mrenna, Maurizio Pierini, Brian Shuve, and Daniel Stolarski for useful conversations. The work of KH and KS is supported by DOE Award DE-SC0015973.

\appendix
\section{Event Simulation of Displaced Simplified Models} 
\label{A}
A strength of the proposed methodology is its reuse of well-established DM simplified models.  In section~\ref{sec:simulation} we considered the case in which new particles ({\it e.g.} $Y^{S}_{0}$, $\chi_{1}$) and interactions are embedded in familiar {\sc DMsimp} models.  Event generation with \mg ~proceeds in exactly the same way as with the original models.  The output LHE contains all of the particle property information needed for {\sc Pythia} to efficiently perform the decays.  The procedure for simulating events in the embedding approach can be summarised as: 

\begin{enumerate}
\item Add the new particle content to the original DM simplified model.  For an EFT decay model, this is simply the new, stable DM particle $\chi_{1}$.  For the simplified decay model the mediating particle must also be included.  
\item Add new interactions to the original model.  These can either be single-parameter EFT operators, or interaction terms involving a mediator.
\item Configure the relevant particle masses and couplings in the \mg ~\texttt{param\_card.dat} to achieve displaced decays. 
\item Generate the $pp \to \chi\bar{\chi}$ in process \mg, which will result in an LHE file that contains the necessary width information in the SLHA header. 
\item Pass the resulting LHE to {\sc Pythia}, which will perform the $\chi \to \chi_{1}X$ decay using the SLHA information.
\end{enumerate}  

This approach is naturally suited to the use of simplified decay models, in which the properties of the mediator and its couplings to the SM and DM sectors are free parameters.  Once these are specified, \mg ~can compute the mediator's partial widths and generate the $\chi\bar{\chi}+{\rm X}$ events. Alternatively, if the widths are calculated by other means ({\it e.g.} analytically from Eqn.~\ref{eq:eff2body}), this information and the new particle content can be input directly to {\sc Pythia}.  In this case, no modifications to the original simplified model are needed.  Event generation with \mg ~requires only that a finite width be assigned to the formerly stable $\chi$.  The simulation procedure in this approach is to:

\begin{enumerate}
\item Configure the $\chi$ mass, width, and SM couplings in the \mg ~\texttt{param\_card.dat} to achieve displaced decays. 
\item Generate the $pp \to \chi\bar{\chi}$ process in \mg. 
\item Pass the resulting the LHE to \texttt{Pythia}.  Any additional new particles ({\it e.g.} $\chi_{1}$), their properties, and their decay modes should be added to the {\sc Pythia} \texttt{ParticleData} table. {\sc Pythia} will then perform the $\chi \to \chi_{1}X$ decay.
\end{enumerate}

Although this approach can be used with both mediated and EFT decay models, it is more simply applied with the latter.  We demonstrate the approach for a t-channel production model with an EFT decay.  This is a concrete example of the GMSB model discussed in section~\ref{sec:GMSB}.  We use the simplified \texttt{SM\_Squark\_udcs\_chi} production model of Ref.~\cite{Papucci:2014iwa}, scan over a range of $\chi$ mass and width values, and generate $\chi\bar{\chi}$ events using \mg.  We then add the $\chi_{1}$ particle to the {\sc Pythia} \texttt{ParticleData} table, configure its mass, and specify a $\chi \to \chi_{1}\gamma$ branching ratio.  Finally, we pass the LHE to {\sc Pythia} to decay the $\chi\bar{\chi}$.  

The left panel of Figure~\ref{fig:GMSBdist} shows the transverse impact parameter ($d_{xy}$) of $\chi_{1}\gamma$ verticies for a range of $\Gamma_{\chi}$ values.  The widths can be directly translated to the SUSY breaking scale in Eqn.~\ref{eq:GMSBwidth}.  The right panel shows distributions of the transverse momentum of the $\chi_{1}\bar{\chi_{1}}$ system, $p_{\rm T}(\chi_{1}\bar{\chi_{1}})$, which approximates the detector-level $\MET$ observable.  The $p_{\rm T}(\chi_{1}\bar{\chi_{1}})$ distributions broaden as the $\chi-\chi_{1}$ mass splitting grows.

\begin{figure}[htbp]
  \centering
  \includegraphics[width=0.48\textwidth]{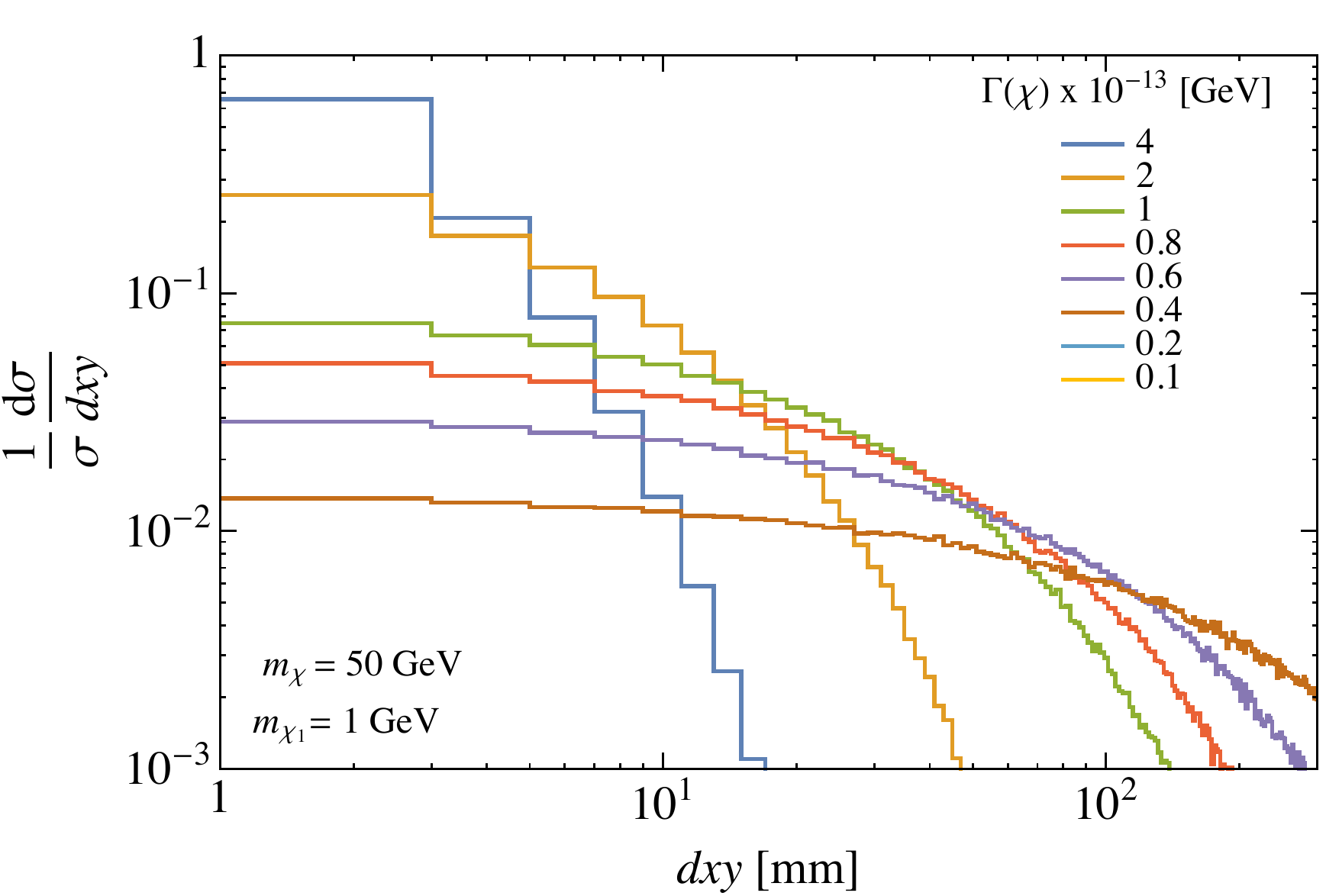}
  \includegraphics[width=0.48\textwidth]{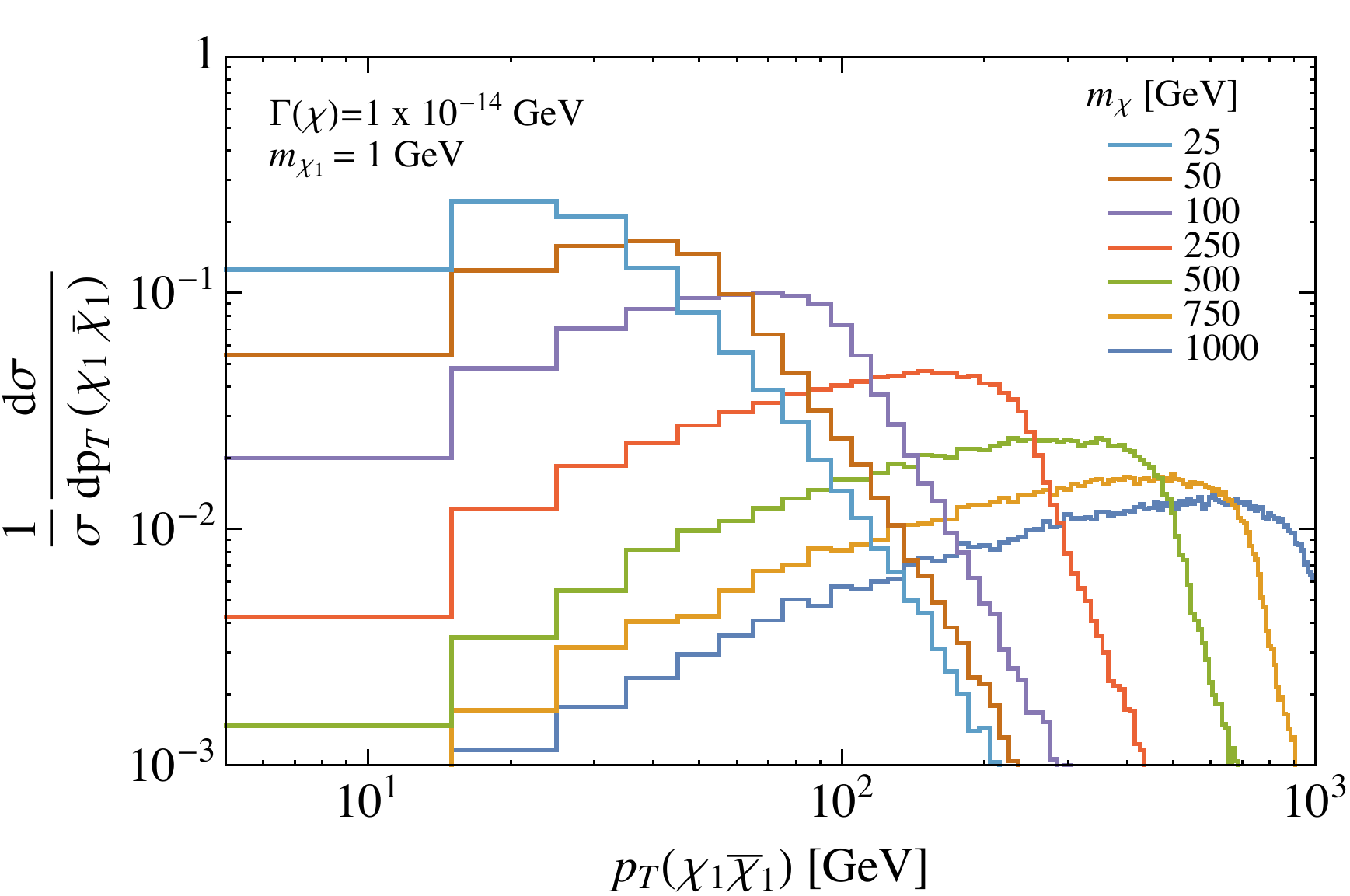}
  \caption{Left: the transverse impact parameter of $\chi_{1}\gamma$ vertices for a range of $\chi$ widths.  Right: the transverse momentum of the DM system ($p_{\rm T}(\chi_{1}\bar{\chi_{1}})$) for various $\chi$ masses.  Other parameters in the GMSB model are fixed as per the panel headings.  The distributions in both panels are unit-normalised. }
  \label{fig:GMSBdist}
\end{figure}

\bibliographystyle{jhep}
\bibliography{displaced}
\end{document}